\documentclass{optica-article}

\journal{opticajournal} 

\articletype{Research Article}

\usepackage{lineno}

\usepackage{glossaries}
\usepackage{braket}
\usepackage{comment}
\usepackage{caption}
\usepackage{subcaption}
\usepackage{placeins}

\newacronym{spdc}{SPDC}{Spontaneous Parametric Down Conversion}
\newacronym{jsa}{JSA}{Joint Spectral Amplitude}
\newacronym{jsi}{JSI}{Joint Spectral Intensity}
\newacronym{jta}{JTA}{Joint Temporal Amplitude}
\newacronym{jti}{JTI}{Joint Temporal Intensity}
\newacronym{umzi}{UMZI}{Unbalanced Mach-Zehnder Interferometer}
\newacronym{bbm92}{BBM92}{Bennett-Brassard-Mermin 1992}
\newacronym{skr}{SKR}{Secret Key Rate}
\newacronym{qber}{QBER}{Quantum Bit Error Rate}
\newacronym{ogs}{OGS}{Optical Ground Station}
\newacronym{fwhm}{FWHM}{Full Width at Half Maximum}
\newacronym{qkd}{QKD}{Quantum Key Distribution}
\newacronym{leo}{LEO}{Low Earth Orbit}
\newacronym{sso}{SSO}{Sun-Synchronous Orbit}
\newacronym{cnes}{CNES}{National Center for Space Studies (Centre National d’Études Spatiales)}
\newacronym{cic}{CIC}{Centre d’Ingénierie Concourante}

\begin{document}

\title{Doppler effect on dual-downlink time-bin quantum key distribution and a proposed ephemeris-assisted witness-field correction}

\author{
Sacha Gressani\authormark{1,2,4*}, 
Luca Furfaro\authormark{2}, 
Erik Kerstel\authormark{3,4},
Jean-Marc Merolla\authormark{2}}

\address{
\authormark{1}AUREA Technology SAS, 6 Rue Sophie Germain, 25000 Besançon, France\\
\authormark{2}Institut FEMTO-ST (UMR 6174), CNRS, Université Marie et Louis Pasteur, F-25000 Besançon, France\\
\authormark{3}Université Grenoble Alpes, Laboratoire Interdisciplinaire de Physique (UMR 5588), 140 Rue de la Physique, 38400 Saint-Martin-d’Hères, France\\
\authormark{4}Centre Spatial Universitaire de Grenoble, 120 Rue de la Piscine, 38400 Saint-Martin-d’Hères, France}

\email{\authormark{*}sacha.gressani@femto-st.fr} 

\begin{abstract*} 
Time-bin entanglement is promising for satellite \gls{qkd} due to its compatibility with telecom fiber networks, but its phase-sensitive measurements are vulnerable to Doppler effects induced by satellite motion. We present a spectral-temporal framework for time-bin-entangled photon pairs generated onboard a satellite via \gls{spdc} and distributed to two \glspl{ogs} over independent downlinks with distinct Doppler shifts. Each \gls{ogs} is equipped with an \gls{umzi} used to perform \gls{bbm92} entangled-photon \gls{qkd}. In a simulated \gls{leo} pass with matched 400~ps analyzer delays, the received time-bin separation changes by at most 8.8~fs per link, while the combined carrier phase shifts by about 38~rad, which primarily degrades phase-basis correlations. After zero-Doppler calibration, this phase separates into independent signal- and idler-link contributions, allowing each station to correct its link Doppler phase shift locally, without real-time phase exchange between the stations. We propose an ephemeris-assisted feed-forward architecture, in which an auxiliary interferometer analyzes a co-propagating classical witness field to reconstruct the local Doppler-induced phase, which a phase modulator then compensates on the quantum channel. In a noise-free model, a 1~Hz sample-and-hold correction rate keeps the phase-basis \gls{qber} below 2\% and retains 93\% of the ideal asymptotic key yield. After 18 accumulated passes, the corrected finite secret key reaches 95\% of its ideal length, whereas relabeling the parity from the ephemeris without optical correction reaches only about 8\%.
\end{abstract*}



\glsresetall

\section{Introduction}
\label{section:Introduction}

\gls{qkd} exploits the principles of quantum mechanics to enable two remote users to establish shared secret keys~\cite{Bennett_2014, Ekert_1991}. In contrast to conventional methods that rely on computational assumptions, \gls{qkd} can provide security without assuming limited computational power of the eavesdropper~\cite{Bennett_2014, Ekert_1991, Scarani_2009}. Extending \gls{qkd} to continental and intercontinental distances remains challenging and has motivated active investigation in both fiber-based and free-space architectures. Within optical fibers, channel transmittance decreases exponentially with distance, which severely limits direct point-to-point transmission without trusted nodes or quantum repeaters~\cite{Pirandola_2017}. Terrestrial free-space links are constrained by atmospheric extinction, line-of-sight availability, and ultimately by Earth's curvature, even though geometric loss increases quadratically with distance~\cite{Kerstel_2018}. Introducing a satellite segment overcomes terrestrial line-of-sight and curvature limitations, while most propagation occurs through the low-loss vacuum of space~\cite{Buttler_1998, Hughes_2000, Rarity_2002, Sidhu_2021, Lu_2022}.

In an entanglement-based satellite architecture, an onboard entangled photon source transmits pairs of entangled signal and idler photons to two distinct \glspl{ogs}. Utilizing the \gls{bbm92} protocol~\cite{BBM92}, Yin et al. demonstrated the feasibility of this approach by distributing polarization-entangled photon pairs from the Micius satellite to two \glspl{ogs} separated by 1203~km, via two independent downlinks~\cite{Yin_2017}. The same platform was later used to demonstrate entanglement-based \gls{qkd} between ground stations separated by 1120~km without a trusted relay~\cite{Yin_2020}.

Energy-time encoding \cite{Franson_1989}, and in particular its discrete analog time-bin encoding, introduced by Brendel et al.~\cite{Brendel_1999}, offers an alternative to polarization encoding by using temporally separated optical modes and time-resolved interferometric measurements. Its compatibility with mature telecommunications components and fiber-optic technologies, as well as its scalability to high-dimensional states \cite{Yu_2025}, makes it well suited for quantum-based networking~\cite{Singh_2026}. Tittel et al.~\cite{Tittel_2000} demonstrated time-bin entanglement-based \gls{qkd}, and Fitzke et al.~\cite{Fitzke_2022} later deployed it in a multi-user optical-fiber network. Telecom-band entangled-photon sources intended for satellite-based \gls{qkd} have also been developed utilizing standard optical-fiber telecommunications components~\cite{Merolla_2023}. Further investigations have demonstrated this approach over turbulent and depolarizing free-space channels~\cite{Jin_2019, Cocchi_2025, Qiu_2025}, enabling interoperability between terrestrial fiber networks and free-space links and encouraging expansion to satellite links~\cite{Cocchi_2025}. However, during transmission, propagation effects and transmitter motion can alter the quantum state's phase and spectral-temporal structure.

Different groups have investigated the effects of propagation and relative motion on entangled-photon-pair states both theoretically and experimentally. MacLean et al.~\cite{MacLean_PRL_2018} directly characterized energy-time entangled photon pairs in both frequency and arrival time, demonstrating that independently controlled dispersion reshapes their joint temporal correlations. They later provided a spectral-temporal description of the Franson interferometric scheme~\cite{Franson_1989}, where the signal and idler interferometers act independently on the \gls{jsa} of the entangled photon pair~\cite{MacLean_PRA_2018}. Yu et al.~\cite{Yu_2025} have demonstrated that dispersion can broaden temporal modes to the point that neighboring coincidence slots are no longer well resolved, degrading interference visibility, increasing the \gls{qber}, and reducing secret-key generation. For moving optical links, Schlake et al.~\cite{Schlake_2025_EPJQT, Schlake_2026} have shown that Doppler transformation of a finite-bandwidth optical wave packet rescales its spectral amplitude and inversely scales its temporal envelope. For temporally separated wave packets, satellite motion causes a Doppler-induced, time-dependent relative phase, as observed by Vallone et al. through single-photon interference between time bins reflected by satellite corner-cube retroreflectors~\cite{Vallone_2016}. In time-bin \gls{qkd}, this phase shift increases the \gls{qber} and reduces the achievable \gls{skr}. Consequently, phase-sensitive measurements require compensation for Doppler-induced phase shifts to maintain quantum state integrity. Multiple methods have been explored to mitigate phase variations and preserve measurement fidelity. 

One approach is the predictive compensation of the Doppler effect based on the orbital ephemeris, as used in low-Earth-orbit radio terminals~\cite{Zantou_2005} and in optical frequency transfer~\cite{Dassie_2023}. In a time-bin system, this predicted Doppler factor can be converted, using the corresponding phase relation, into the required phase correction. However, because this approach relies on a modeled orbital Doppler evolution rather than an optical measurement of the received phase, its accuracy is set by that of the ephemeris, and it does not, by itself, account for additional phase perturbations in the optical system, such as those caused by pump frequency drift, thermal fluctuations, or mechanical vibrations of the source or of the receiver interferometers~\cite{Chapman_2020, Chapman_2022, Qiu_2025}. 

A second approach involves monitoring the phase shift in real time and applying active correction to compensate for its temporal evolution. Chapman et al. demonstrated active compensation by emulating satellite-induced Doppler shifts in a single-link polarization–time-bin system utilizing a classical optical reference and a feedback-controlled analyzer interferometer, first for superdense teleportation~\cite{Chapman_2020} and later for \gls{qkd}~\cite{Chapman_2022}. Outside the satellite context, Tagliavacche et al. used a co-propagating classical control field in entanglement-based frequency-bin \gls{qkd} to track phase drift between frequency bins caused by optical fiber path variations and adaptively rotate the quantum measurement basis using electro-optic phase modulation~\cite{Tagliavacche_2025}. In principle, such optical-reference schemes can compensate for any phase perturbation shared by the reference field and the quantum signal, so that their effectiveness depends on the optical paths and components common to both fields, as proposed by Jiao et al. for classical coherent satellite-to-ground optical links~\cite{Jiao_2026}.

Alternatively, the protocol can be designed to tolerate phase evolution rather than actively compensate for it. Tannous et al. demonstrated a reference-frame-independent time-bin \gls{qkd} scheme suitable for moving free-space links, in which sufficiently slow Doppler-induced phase variations can be tolerated~\cite{Tannous_2025}.

Previous studies have examined how satellite motion affects frequency-bin encoding. Vinet et al. modeled Doppler-induced phase evolution along a single free-space link and estimated the phase-compensation bandwidth below 1~Hz for both frequency-bin and time-bin encoding in a low-Earth-orbit scenario~\cite{Vinet_2025}. More recently, the same group demonstrated passive time-resolved measurement of frequency-bin entanglement and proposed its extension to reference-frame-independent \gls{qkd} within a dual-downlink satellite configuration~\cite{Vinet_2026}.

\begin{figure}[htbp]
    \centering
    \includegraphics[width=.7\linewidth]{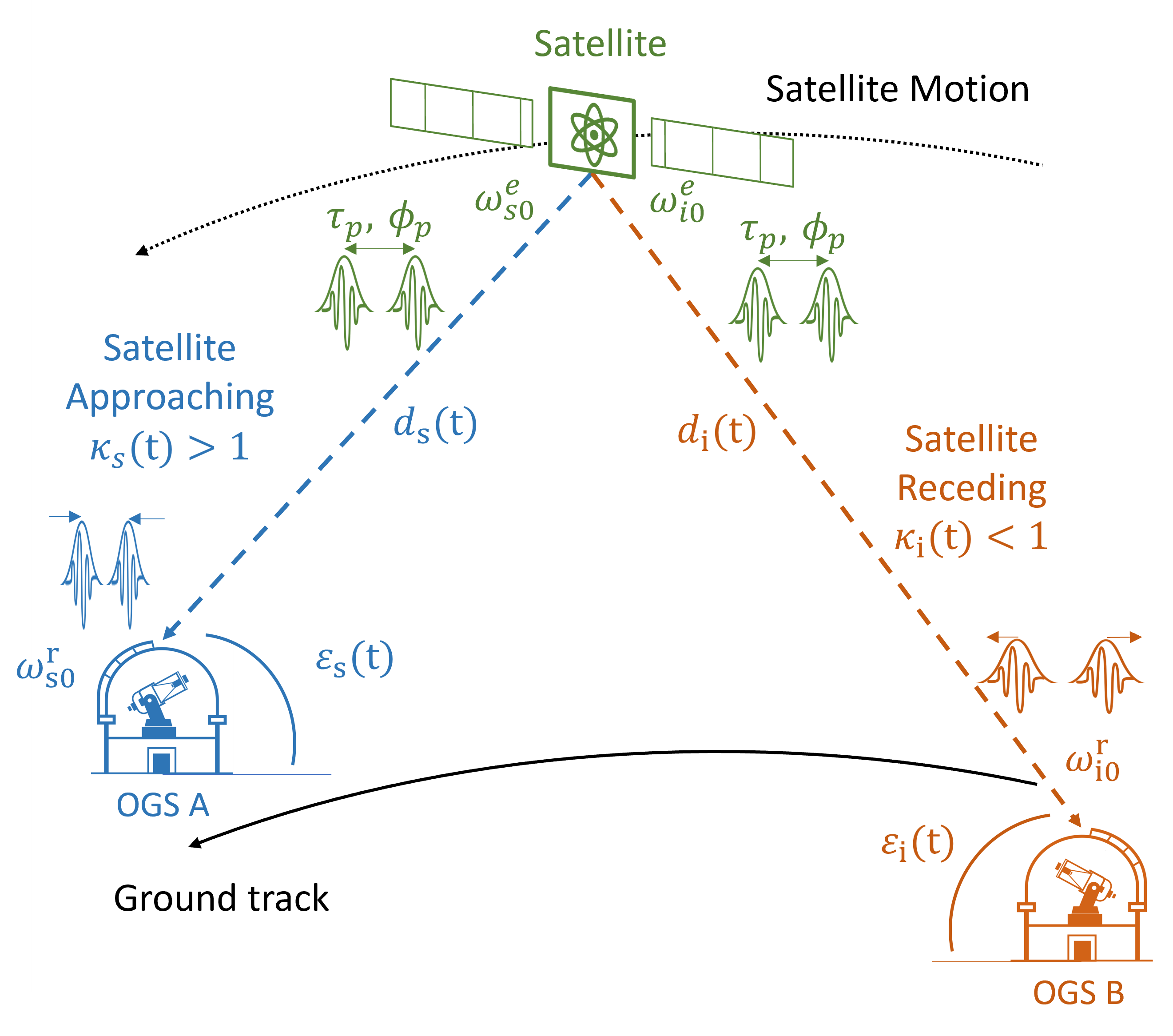}
    \caption{Dual-downlink geometry and Doppler transformation of the distributed time-bin state. An entangled-photon source onboard the satellite generates signal (s, blue) and idler (i, orange) photon pairs which are distributed to two spatially separated \glspl{ogs} over independent downlinks, indexed by $j\in\{s,\, i\}$. The emitted photon pair is prepared in a coherent superposition of early and late time bins separated by a temporal interval $\tau_p$, and characterized by a relative phase $\phi_p$. The signal and idler photon carriers have angular frequencies $\omega_{s0}^{e}$ and $\omega_{i0}^{e}$, respectively. Each downlink is characterized by its time-dependent slant range $d_j(t)$, Doppler factor $\kappa_j(t)$, and elevation angle $\varepsilon_j(t)$. For an approaching signal link, $\kappa_s(t) > 1$, while for a receding idler link, $\kappa_i(t) < 1$. Consequently, the received carrier frequencies are blue- and red-shifted, respectively, and the associated time-bin separations compress and expand. Both effects are illustrated schematically for clarity and are not drawn to scale.}
    \label{fig:Dual-downlink QKD}
\end{figure}

These investigations establish several approaches to phase stabilization and phase-tolerant quantum communication. However, these works do not investigate how the Doppler effect affects two-photon interference when time-bin-entangled photons, produced by \gls{spdc}, experience independent Doppler shifts in the downlinks. In this work, we analyze the influence of Doppler-induced effects on phase-sensitive \gls{bbm92} observables and investigate which resulting phase contributions the two receivers can compensate for. We extend the spectral-temporal interferometric approach implemented by MacLean et al. for energy-time entanglement~\cite{MacLean_PRA_2018} to time-bin-entangled photon pairs distributed from a satellite to two independent \glspl{ogs}. Fig.~\ref{fig:Dual-downlink QKD} summarizes the dual-downlink configuration considered here.

Using the Doppler transformation of Schlake et al.~\cite{Schlake_2025_EPJQT, Schlake_2026}, we derive the received biphoton state for distinct signal- and idler-link Doppler factors. We examine spectral and temporal rescaling via joint spectral and temporal intensities, both before and after propagation through receiver interferometers, and evaluate its effect on two-photon interference. We then derive the central-slot interference phase, isolating the Doppler-induced carrier contribution from the detuning-dependent terms. For matched analyzer delays and the specified zero-Doppler calibration, we show that the carrier-phase contribution separates into independent signal- and idler-link components, allowing local correction at each \gls{ogs}. To estimate the relevant Doppler effects under realistic link dynamics, we apply the model to a selected \gls{leo} dual-downlink pass. We first distinguish the carrier-phase and detuning-dependent Doppler contributions and evaluate the sensitivity to analyzer-delay mismatch. From the resulting carrier-phase evolution, we then determine the required phase-correction rate. To implement this correction locally, we propose an active, ephemeris-assisted feed-forward Doppler correction architecture that operates independently at each \gls{ogs}. At each receiver, we analyze a co-propagating, wavelength-multiplexed classical witness field with a known emitted carrier frequency using an auxiliary \gls{umzi} calibrated at the witness frequency. Each \gls{ogs} combines the measured contrast with ephemeris information to reconstruct the local witness phase, then converts it to the corresponding quantum-channel phase and compensates it with a time-bin-selective phase modulator. Since the witness shares only the free-space downlink with the quantum signal, it tracks the link-dependent Doppler phase but not the drift of the pump or of the analyzer interferometers. The two downlinks are corrected independently, without real-time phase exchange between the two \glspl{ogs}. Finally, we evaluate the resulting \gls{bbm92} performance with and without optical correction, in both the asymptotic and finite-key regimes, including the accumulation of detections over successive passes.

\section{Theoretical framework}
\label{section:Theoretical framework}

This work builds on the source-and-receiver architecture of the experimental energy-time (time-bin) \gls{qkd} system demonstrated by Merolla et al., which uses a continuous-wave-pumped type-0 \gls{spdc} source, fiber-based \gls{umzi}, and passive basis selection for \gls{bbm92}~\cite{Merolla_2023}. We adapt this architecture to a pulsed-pump time-bin implementation and extend it from the static, attenuator-emulated combined downlink loss to a dynamic dual-downlink satellite scenario, analyzing the effects of independent Doppler-induced phase shifts and their local compensation.

\subsection{Time-bin entangled photon source}
\label{subsection:Time-bin entangled photon source}

\begin{figure}[htbp]
    \centering
    \includegraphics[width=1\linewidth]{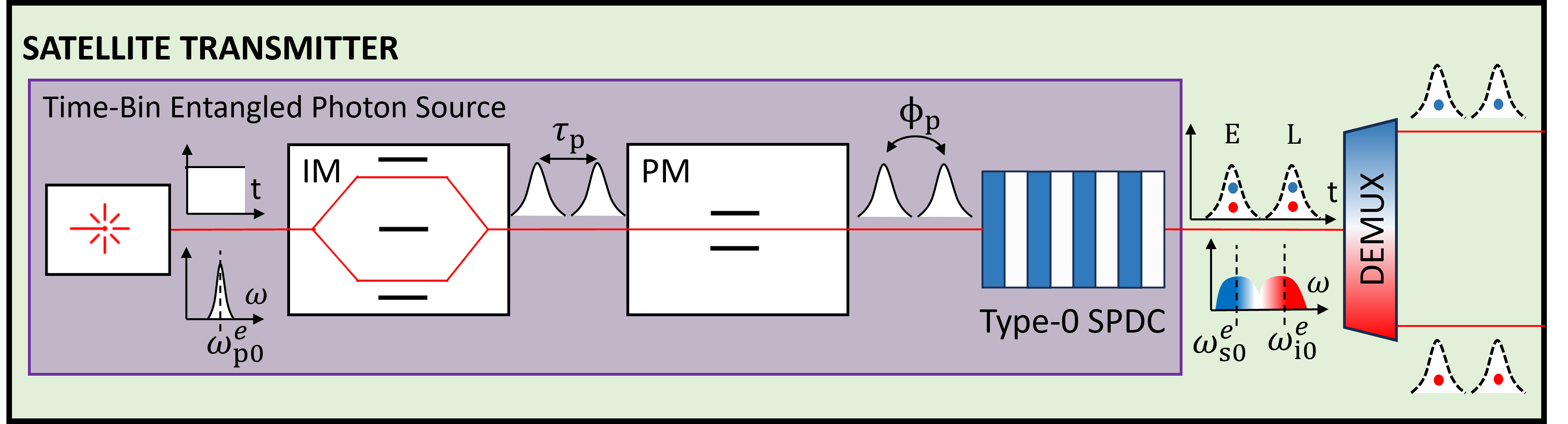}
    \caption{Time-bin entangled-photon source architecture. An intensity modulator carves a continuous, narrow-linewidth pump field with carrier angular frequency $\omega_{p0}^e$ into two pulses of equal amplitude, denoted as early (E) and late (L), separated by a temporal delay $\tau_p$. A phase modulator sets a well-defined relative phase, $\phi_p$, between the two pump pulses, which are then injected into a type-0 nonlinear crystal, generating co-polarized, nondegenerate signal (blue) and idler (red) photon pairs through \gls{spdc}. These photon pairs possess carrier angular frequencies $\omega_{s0}^e$ and $\omega_{i0}^e$, respectively, and satisfy the energy conservation relation $\omega_{p0}^e = \omega_{s0}^e + \omega_{i0}^e$. The pair-creation process preserves the pump's temporal coherence, producing signal–idler photon pairs in a coherent superposition of early- and late-time bins. The dashed outlines of the pulses indicate the quantum uncertainty associated with whether the early or late pump pulse produces the pair. A demultiplexer subsequently separates the signal and idler wavelength channels.}
    \label{fig:Time-bin entangled-photon source}
\end{figure}

Fig.~\ref{fig:Time-bin entangled-photon source} presents the architecture of the time-bin entangled-photon source onboard the satellite. The time-bin separation $\tau_p$ is chosen to be much longer than the pump pulse duration but much shorter than the pump field's coherence time. This choice minimizes temporal overlap and guarantees a well-defined relative phase $\phi_p$ between the early and late pump pulses. 

In the low-gain regime, the emitted time-bin state is
\begin{equation}
\begin{aligned}
\label{equation:Emitted Time-Bin State}
\ket{\Psi^e}
&=
\frac{1}{\sqrt{2}}
\int d\omega_s^ed\omega_i^e
F^e(\omega_s^e,\omega_i^e)
\left[
1 +
e^{i\phi_p}
e^{i(\omega_s^e + \omega_i^e)\tau_p}
\right]
a_s^{e\dagger}(\omega_s^e)
a_i^{e\dagger}(\omega_i^e)
\ket{0},
\end{aligned}
\end{equation}
where $1/\sqrt{2}$ is a normalizing factor to preserve unit area of $\|\Psi^e\|^2$. The superscript $e$ indicates quantities in the emitter frame. For the signal and idler modes, indexed by $j \in \{s,i\}$, $a_j^{e\dagger}(\omega)$ denotes the photon creation operator at angular frequency $\omega$. Together with the destruction operator $a_j^{e}(\omega)$ they satisfy the commutation relation $[a_j^e(\omega), a_k^{e\dagger}(\omega')] = \delta_{jk}\delta(\omega - \omega')$. $F^e(\omega_s^e,\omega_i^e)$ represents the single-pulse \gls{jsa}. The \gls{jta} is obtained by taking the symmetric two-dimensional Fourier transform of the single-pulse \gls{jsa}
\begin{equation}
\label{eq:jta}
f^e(t_s^e,t_i^e)
=
\frac{1}{2\pi}
\int d\omega_s^ed\omega_i^e
F^e(\omega_s^e,\omega_i^e)
e^{-i\omega_s^et_s^e}
e^{-i\omega_i^et_i^e},
\end{equation}
where $t_s^e$ and $t_i^e$ are the temporal coordinates conjugate to the emitted signal and idler frequencies, respectively. $\|f^e(t_s^e,t_i^e)\|^2$ corresponds to the single-pulse joint temporal intensity. With this convention, Parseval's theorem yields
\begin{equation}
\label{eq:normalization}
\int d\omega_s^ed\omega_i^e
\left|
F^e(\omega_s^e,\omega_i^e)
\right|^2
=
\int dt_s^edt_i^e
\left|
f^e(t_s^e,t_i^e)
\right|^2
= 1.
\end{equation}

Under the broadband phase-matching approximation, the single-pulse \gls{jsa} factorizes into the product of the pump spectral shape function, the filtered signal, and the filtered idler envelope functions~\cite{Yu_2025}. We model the \gls{jsa} functions as a Gaussian function given by
\begin{equation}
\begin{aligned}
\label{equation:Gaussian JSA}
F^e(\omega_s^e,\omega_i^e) 
&=
\mathcal{N}
\exp{
\left\{
-\frac{(\omega_s^e + \omega_i^e - \omega_{p0}^e)^2}{2\sigma_{\omega, p}^2}
\right\}
}
\exp{
\left\{
-\frac{(\omega_s^e - \omega_{s0}^e)^2}{2\sigma_{\omega,s}^2}
\right\}
}
\exp{
\left\{
-\frac{(\omega_i^e - \omega_{i0}^e)^2}{2\sigma_{\omega,i}^2}
\right\}
},
\end{aligned}
\end{equation}
where $\mathcal{N}$ denotes the normalization constant determined by Eq.~\eqref{eq:normalization}. Each emitted frequency is expressed as $\omega_j^e = \omega_{j0}^e + \Omega_j^e$, where $\omega_{j0}^e$ represents the emitted carrier frequency, and $\Omega_j^e$ is the frequency detuning for mode $j \in \{s,i\}$. Similarly, the pump frequency can be written as $\omega_p^e = \omega_{p0}^e + \Omega_p^e$, where $\omega_{p0}^e$ corresponds to the pump carrier frequency and $\Omega_p^e$ is the frequency detuning. The carrier frequencies satisfy the energy conservation relation $\omega_{p0}^e = \omega_{s0}^e + \omega_{i0}^e$. Energy conservation for each pump spectral component, $\omega_p^e = \omega_s^e + \omega_i^e$, implies $\Omega_p^e = \Omega_s^e + \Omega_i^e$. The parameters $\sigma_{\omega, p}$, $\sigma_{\omega, s}$, and $\sigma_{\omega, i}$ are the standard deviations of the Gaussian field-amplitude functions, representing their $1/\sqrt{e}$ half-widths. For a Gaussian amplitude with standard deviation $\sigma_x$, the intensity standard deviation is $\sigma_x^I = \sigma_x/\sqrt{2}$ for $x \in \{\omega,t\}$, and the intensity \gls{fwhm} is $\Delta x_{\mathrm{FWHM}}^I = 2\sqrt{2\ln(2)}\sigma_x^I$. Using the Fourier-transform convention of Eq.~\eqref{eq:jta}, $\sigma_t\,\sigma_\omega = 1$. 

Throughout this analysis, the pump carrier frequency $\omega_{p0}^e$, time-bin separation delay $\tau_p$, and relative pump phase $\phi_p$ are assumed to be constant during the satellite pass. We focus on the narrow-pump regime $\sigma_{\omega, p} \ll \sigma_{\omega,s}, \sigma_{\omega,i}$, where the joint spectrum exhibits strong frequency anticorrelation. For the parameters considered here, $\sigma_{\omega,p}^{I} \simeq 1.2\times10^{10}~\mathrm{rad \cdot s^{-1}}$, and the signal and idler bandwidths are at least $27$ times larger.

\begin{table}[htbp]
\centering
\caption{Spectral and temporal parameters of the time-bin-entangled photon source used in the numerical analysis.}
\label{tab:source_params}
\begin{tabular}{lll}
\hline
\textbf{Symbol} & \textbf{Value} & \textbf{Description} \\
\hline
$\lambda_s$ & 1520~nm & Signal central wavelength \\
$\lambda_i$ & 1580~nm & Idler central wavelength \\
$\tau_p$ & 400~ps & Pump time-bin separation \\
$\Delta t_{FWHM, p}^I$ & 100~ps & Pump intensity FWHM \\
$\Delta\lambda_{FWHM, s, i}^I$ & 1~nm & Signal/idler filter intensity FWHM\\
\hline
\end{tabular}
\end{table}

\begin{figure}[htbp]
    \centering
    \includegraphics[
        width=.49\columnwidth
    ]{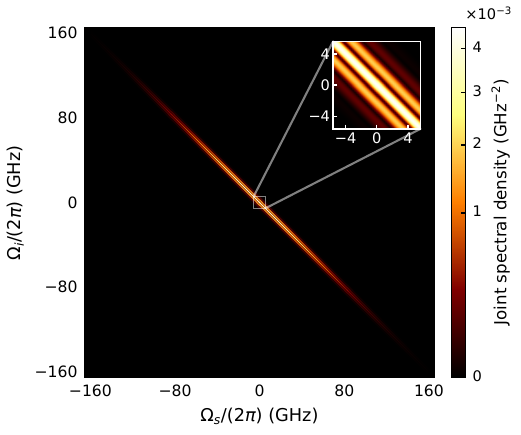}
    \includegraphics[
        width=.49\columnwidth
    ]{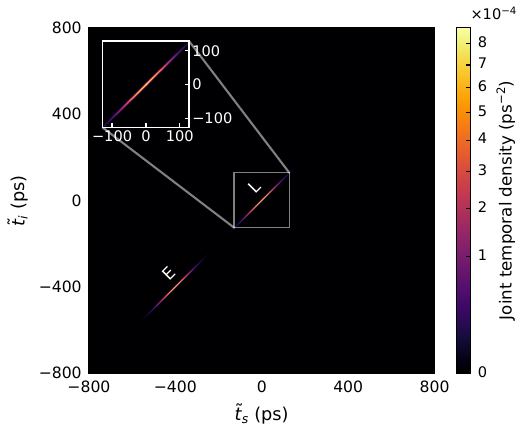}
    \caption{Joint spectral (\textbf{left}) and temporal (\textbf{right}) intensities of the emitted time-bin state. The joint spectral intensity is plotted as a function of the signal and idler frequency detunings following the energy-conserving antidiagonal $\Omega_s^e + \Omega_i^e \simeq 0$ and exhibits parallel fringes arising from the coherent superposition of early and late pair-generation amplitudes. The box defines the area magnified in the inset, where $|\Omega_{s, i}^e| \leq 6\,\sigma^I_{\omega, p}$. The corresponding joint temporal intensity, displayed in the shifted emitter-frame coordinates $\tilde{t}_s^e$ and $\tilde{t}_i^e$, shows the early and late wave packets recentered at $(-\tau_p,-\tau_p)$ and (0, 0), respectively, along the diagonal $\tilde t_s^e = \tilde t_i^e$. The box delimits the region magnified in the inset, where $|\tilde{t}_s^e|,|\tilde{t}_i^e| \leq 6\,\sigma_{t,p}^I$.}
    \label{fig:Source Joint Intensities}
\end{figure}

Figure~\ref{fig:Source Joint Intensities} depicts the spectral and temporal correlation characteristics of the emitted time-bin state using the parameters listed in Table~\ref{tab:source_params}. The joint spectral intensity is concentrated along the energy-conserving antidiagonal $\Omega_s^e + \Omega_i^e \simeq 0$, indicating strong frequency anticorrelation. The coherent superposition of early and late pair-generation amplitudes produces parallel spectral fringes. To analyze the joint temporal intensity, we defined shifted emitter-frame temporal coordinates $\tilde{t}_j^e = t_j^e-\tau_p$, with $j\in\{s, i\}$, such that the late pair-generation amplitude is centered at the origin. Consequently, the joint temporal intensity features two pulses corresponding to early and late pair generation, centered at $(\tilde{t}_s^e,\tilde{t}_i^e) = (-\tau_p,-\tau_p)$ and (0, 0), respectively.

\subsection{Doppler transformation of the time-bin state}
\label{subsection:Doppler transformation of the time-bin state}

As illustrated in Fig.~\ref{fig:Dual-downlink QKD}, the Doppler effect shifts the satellite's emitted frequencies relative to those received by the \glspl{ogs}. For each downlink $j\in\{s, i\}$, we use the longitudinal Doppler factor $\beta_j =  -\dot{d}_j/c$ to model the Doppler stretching factor
\begin{equation}
\label{equation:Doppler Factor}
\kappa_j
=  
\frac{\omega^r_j}{\omega^e_j}
= 
\sqrt{\frac{1 + \beta_j}{1 - \beta_j}}.
\end{equation}
Here, $\dot{d}_j = \mathbf{v}_{SR, j} \cdot \mathbf{u}_{SR, j}$ defines the slant-range rate, with $\mathbf{v}_{SR, j} = \mathbf{v}_{R, j} - \mathbf{v}_S$ representing the velocity of receiver $j$ relative to the source, and $\mathbf{u}_{SR, j}$ is the unit vector directed from the source to the receiver. According to the sign convention in Eq.~\eqref{equation:Doppler Factor}, an approaching receiver is characterized by $\dot{d}_j < 0$, which corresponds to $\beta_j > 0$ and $\kappa_j > 1$. Consequently, the received optical frequency is greater than the emitted frequency, resulting in a blueshift. Both the biphoton coherence time and the time-bin separation $\tau_p$ are assumed to be much shorter than the timescale of orbital evolution, allowing the line-of-sight velocity, and thus $\kappa_j$, to be treated as constant during the time-bin separation. The validity of this quasistatic approximation is confirmed a posteriori in Sec.~\ref{subsection:Doppler-factor and carrier-phase evolution}.

Applying the frequency rescaling $\omega^r_j = \kappa_j\omega^e_j$ independently to the signal and idler modes, the received time-bin-entangled state becomes
\begin{equation}
\label{equation:Received Time-Bin State}
\begin{aligned}
\ket{\Psi^r}
&=
\frac{1}{\sqrt{2}}
\int d\omega_s^r d\omega_i^r
\frac{1}{\sqrt{\kappa_s\kappa_i}}
F^e
\left(
\frac{\omega_s^r}{\kappa_s},
\frac{\omega_i^r}{\kappa_i}
\right)
\left[
1 +
\exp\left\{
i(\omega_s^r + \omega_i^r)\tau_p
+
i\phi_p
-
i\sum_{j = s,i}\phi_j^D(\omega_j^r)
\right\}
\right]\\
&\quad\times
a_s^\dagger(\omega_s^r)
a_i^\dagger(\omega_i^r)
\ket{0}.
\end{aligned}
\end{equation}
The detailed derivation of the above equation is shown in Appendix~\ref{appendix:Doppler-induced transformation of the time-bin biphoton state}. For downlink $j$, the Doppler effect transforms the emitted time-bin separation $\tau_p$ to $\tau_p/\kappa_j$ in the receiver's frame. Expressing the received  state in terms of the nominal time-bin separation $\tau_p$ consequently imparts a frequency-dependent phase shift,
\begin{equation}
\label{equation:Doppler-induced phase mismatch}
\phi_j^D(\omega_j^r)
=
\omega_j^r\left(
1 - \frac{1}{\kappa_j}
\right)
\tau_p
=
\omega_j^e
\left(
\kappa_j - 1
\right)
\tau_p.
\end{equation}
The received angular frequency is decomposed as $\omega_j^r = \omega_{j0}^r + \Omega_j^r$, where $\omega_{j0}^r = \kappa_j\omega_{j0}^e$, and $\Omega_{j}^r = \kappa_j\Omega_{j}^e$. Substituting $\omega_j^r = \omega_{j0}^r + \Omega_j^r$ into Eq.~\eqref{equation:Doppler-induced phase mismatch} allows the Doppler-induced phase to be separated into carrier and detuning terms,
\begin{equation}
\label{equation:Doppler phase receiver frame}
\phi_j^D (\omega_j^r)
=
\underbrace{\omega_{j0}^r
\left(
1 - \frac{1}{\kappa_j}
\right)
\tau_p}_{\mathrm{carrier}}
+ 
\underbrace{\Omega_j^r
\left(
1 - \frac{1}{\kappa_j}
\right)
\tau_p}_{\mathrm{detuning}}.
\end{equation}
Equivalently, using emitter-frame variables,
\begin{equation}
\label{equation:Doppler phase emitter frame}
\phi_j^D (\omega_j^e)
=
\underbrace{\omega_{j0}^e
(\kappa_j - 1)
\tau_p}_{\mathrm{carrier}}
+ 
\underbrace{\Omega_j^e
(\kappa_j - 1)
\tau_p}_{\mathrm{detuning}}.
\end{equation}
The carrier term shifts the phase-basis origin, whereas the detuning-dependent term can degrade interference visibility, as different photon pairs sampled from the joint spectrum may acquire distinct phases that can partially cancel when integrating over the entire joint spectrum. 

\subsection{Interferometric analysis and BBM92 basis selection}
\label{subsection:Interferometric analysis and BBM92 basis selection}

\begin{figure}[htbp]
    \centering
    \includegraphics[width=.8\linewidth]{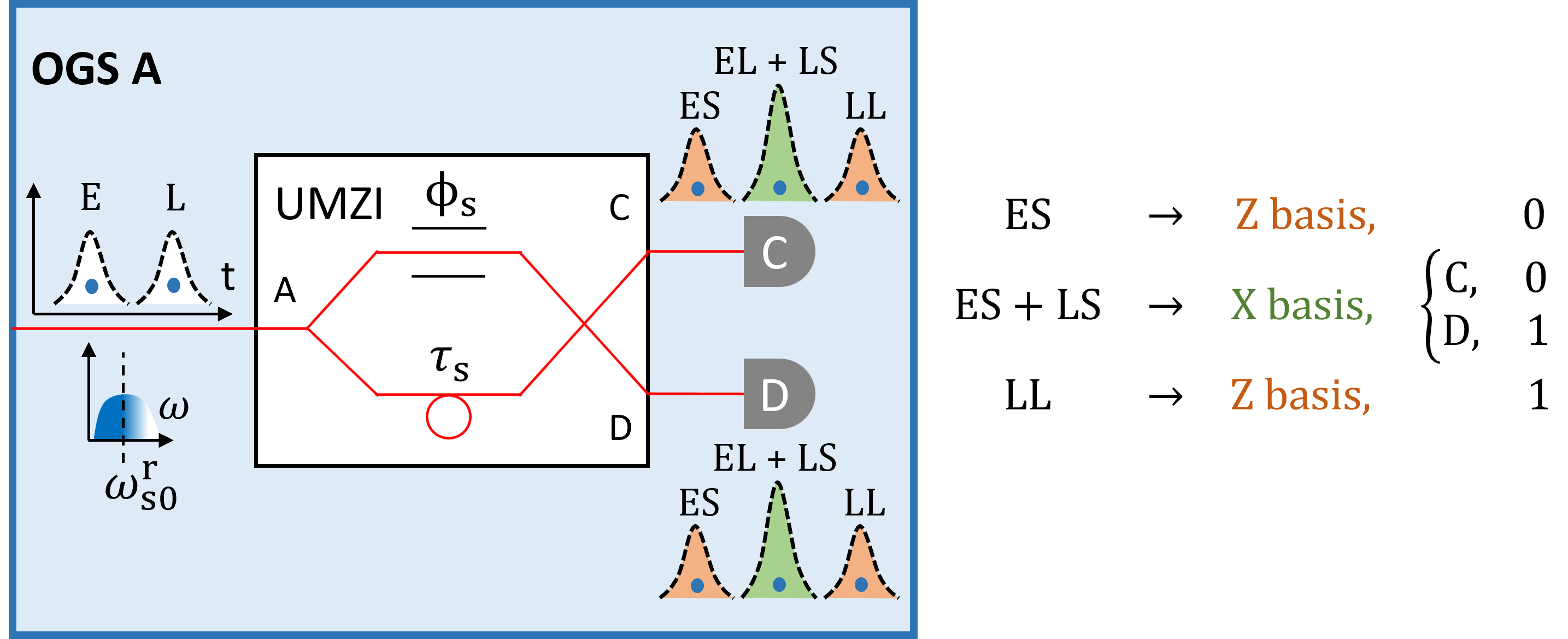}
    \caption{Receiver architecture. The incoming signal is analyzed by an \gls{umzi} which introduces a relative delay $\tau_s$ and a tunable phase $\phi_s$. At each output port ($C$ and $D$), photon detection events are temporally resolved into three distinct arrival-time slots: $ES$, $EL + LS$, and $LL$. The outer slots (orange) correspond to time-basis ($Z$-basis) measurements, while the central slot (green) corresponds to phase-basis ($X$-basis) measurements. The indicated binary values denote the corresponding local measurement outcomes.}
    \label{fig:Signal receiver}
\end{figure}

At each \gls{ogs}, photons received in channel $j \in \{s,i\}$ are analyzed using an \gls{umzi} with a delay $\tau_j$, depicted in Fig.~\ref{fig:Signal receiver}. The \gls{umzi} features a controllable phase $\phi_j$ and two output ports, $C$ and $D$. When the source and analyzer delays are matched, $\tau_p = \tau_j$, the detection times at each output port are distributed across three distinct arrival-time slots.  These slots arise from the possible combinations of the pump time bin and the analyzer path. Each contribution is labeled with a two-letter code: the first letter, $E$ or $L$, denotes generation in the early or late pump time bin, and the second, $S$ or $L$, indicates propagation through the short or long arms of the \gls{umzi}. As explained in Merolla et al.~\cite{Merolla_2023}, the resolved arrival-time slots enable passive \gls{bbm92} basis selection. Detection in either outer slot corresponds to a time-basis ($Z$) measurement, with $ES$ and $LL$ assigned binary outcomes $0$ and $1$, respectively. Detection in the central $EL + LS$ slot corresponds to a phase-basis ($X$) measurement, with output ports $C$ and $D$ assigned outcomes $0$ and $1$, respectively.

Having established the protocol for local time-bin measurement and basis assignment, we now describe the receiver-induced transformation. To this end, we apply the following transformation to the quantum state defined in Eq.~\eqref{equation:Received Time-Bin State}:
\begin{equation}
\begin{aligned}
\label{equation:UMZI Mode Transformation}
a^\dagger(\omega_j^r)
&\rightarrow
\frac{i}{2}
\left[
1 + e^{i(\omega_j^r\tau_j + \phi_j)}
\right]
c^\dagger(\omega_j^r)
+
\frac{1}{2}
\left[
1 - e^{i(\omega_j^r\tau_j + \phi_j)}
\right]
d^\dagger(\omega_j^r),
\end{aligned}
\end{equation}
where $c^\dagger(\omega_j^r)$ and $d^\dagger(\omega_j^r)$ denote the photon creation operators for output ports $C$ and $D$ of analyzer $j$, respectively. For notational clarity in subsequent calculations, we assign port signs $s_C = +1$ and $s_D = -1$ to the $C$ and $D$ outputs, respectively.

Applying Eq.~\eqref{equation:UMZI Mode Transformation} to both photons in the received state, Eq.~\eqref{equation:Received Time-Bin State}, produces eight distinct two-photon path amplitudes, as detailed in Appendix~\ref{appendix:Analyzer path expansion and central-slot projection}. Since the joint coincidence structure is determined by both the pump time bin and the propagation paths through the signal and idler analyzers, we generalize the local two-letter notation from Fig.~\ref{fig:Signal receiver} to a three-letter code. The first letter specifies the pump time bin, and the second and third correspond to the paths taken by the signal and idler photons, respectively. Relative to the early-bin reference, a late pump bin shifts both photon arrival times by $\tau_p$, while traversing the long arm in the signal or idler interferometer imparts an additional delay of $\tau_s$ or $\tau_i$ to the respective photon. To resolve these pathway contributions in the coincidence measurement, we define the mean and relative received-time coordinates as $t_+^r = (t_s^r + t_i^r)/2$ and $t_-^r = t_s^r - t_i^r$, where $t_s^r$ and $t_i^r$ represent the arrival times of the signal and idler photons, respectively. The coordinate $t_+^r$ identifies the shared arrival-time slot, while $t_-^r$ quantifies the temporal offset between the two photons.

\begin{figure}[htbp]
    \centering
    \includegraphics[
        width=.49\columnwidth
    ]{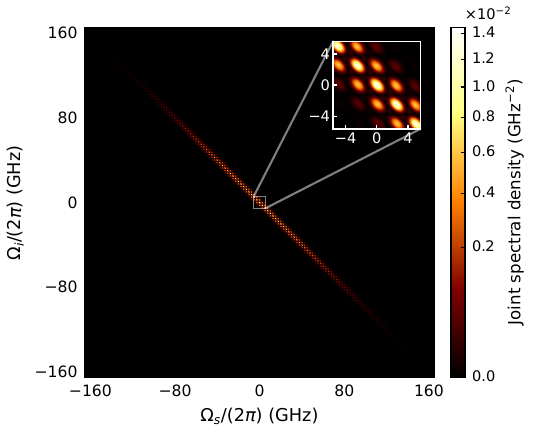}
    \includegraphics[
        width=.49\columnwidth
    ]{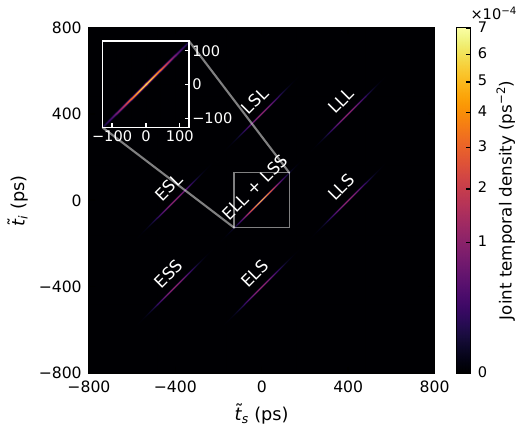}%
    \vspace{-0.4em}    
    \includegraphics[
        width=0.49\columnwidth
    ]{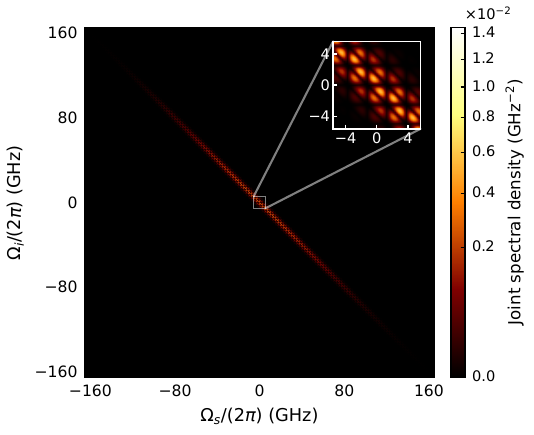}
    \includegraphics[
        width=0.49\columnwidth
    ]{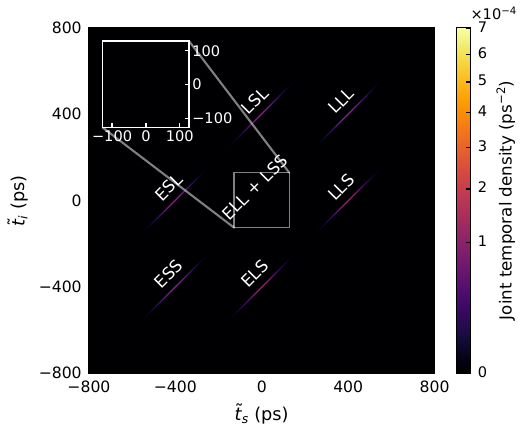}%
\caption{Joint spectral (\textbf{left column}) and temporal (\textbf{right column}) intensities are shown for the $CC$ (\textbf{top}) and $CD$ (\textbf{bottom}) output pairs under the zero-Doppler configuration ($\kappa_s = \kappa_i = 1$), with matched \gls{umzi} delays ($\tau_s = \tau_i = \tau_p$) and the calibrated zero-Doppler phase reference. Joint Spectral and temporal intensities are normalized so that the corresponding $CC$  output pair distributions integrate to unity. The $CD$ output pair distributions use the same normalization and color scale as the corresponding $CC$ distributions, enabling direct comparison between the two output-port pairs. Temporal coordinates are recentered on the $\ket{ELL}/\ket{LSS}$ coincidence slot at $\tilde{t}_s = \tilde{t}_i = 0$. In the temporal domain (\textbf{right column}), the bins align along the constant-relative-delay lines $t_-^r = t_s^r - t_i^r = 0, \pm \tau_p$. At the $CC$ output (\textbf{top right}), the $\ket{ELL}$ and $\ket{LSS}$ amplitudes interfere constructively, producing a bright central bin. At the $CD$ output (\textbf{bottom right}), the idler $D$ port introduces a relative $\pi$ phase, resulting in destructive interference and suppression of the central bin. For both outputs, this phase change does not affect the six noncentral bin intensities. In the spectral domain (\textbf{left column}), the distribution is localized near the energy-conserving antidiagonal $\Omega_s^r + \Omega_i^r \simeq 0$. The spectral insets display a two-dimensional interference pattern arising from the coherent superposition of time bins produced by the two \glspl{umzi}. This time-bin superposition induces a periodic modulation along the sum-frequency coordinate, $\nu_+ = (\Omega_s^r + \Omega_i^r)/(2\pi)$.}
\label{fig:Joint intensities}
\end{figure}

In the zero-Doppler configuration, where $\kappa_s = \kappa_i = 1$ and $\tau_s = \tau_i = \tau_p$, the eight two-photon path amplitudes are mapped onto seven distinct coincidence bins. This is due to the overlap of the $\ket{ELL}$ and $\ket{LSS}$ pathways at $(t_+^r, t_-^r) = (\tau_p, 0)$, while the remaining six pathways retain temporal distinguishability. Quantum interference between these two temporally indistinguishable amplitudes implements the phase ($X$) basis measurement of the \gls{bbm92} protocol, with analyzer output ports encoding the binary outcomes. The noninterfering $\ket{ESS}$ and $\ket{LLL}$ amplitudes, localized at (0, 0) and $(2\tau_p, 0)$, respectively, correspond to the early and late outcomes of the time-bin ($Z$) basis measurement. The four residual pathways, wherein one photon propagates through the short arm and the other through the long arm of their respective \glspl{umzi}, are systematically excluded during basis reconciliation~\cite{Tittel_2000, Fitzke_2022}. Figure~\ref{fig:Joint intensities} shows the joint spectral and temporal intensities for the $CC$ and $CD$ analyzer-output pairs. We assign the first letter to the signal and the second to the idler. These distributions establish the zero-Doppler reference for subsequent analysis of Doppler-induced phase perturbations detailed in Sec.~\ref{subsection:Doppler contribution to phase-basis correlations}. For analysis of the joint temporal intensities, we recenter the received arrival-time coordinates as $\Tilde{t}_s^r = t_s^r - \tau_p$ and $\Tilde{t}_i^r = t_i^r - \tau_p$, such that the $\ket{ELL}/\ket{LSS}$ interference is centered at $(\Tilde{t}_s^r, \Tilde{t}_i^r) = (0, 0)$. 

By retaining the two temporally indistinguishable $\ket{ELL}$ and $\ket{LSS}$ contributions from the full Doppler-dependent expression detailed in Appendix~\ref{appendix:Analyzer path expansion and central-slot projection}, we define the phase-reduced central-slot state component for output-port pair $\alpha,\beta \in \{C,D\}$ (see Appendix~\ref{appendix:Analyzer path expansion and central-slot projection})
\begin{equation} 
\begin{aligned} 
\label{equation:phase-reduced central-slot state component} 
\ket{\Psi_{\alpha\beta}^{r, X}} 
&=
\frac{1}{4\sqrt{2}}
\int d\omega_s^rd\omega_i^r \frac{1}{\sqrt{\kappa_s\kappa_i}} F^e 
\left(
\frac{\omega_s^r}{\kappa_s}, \frac{\omega_i^r}{\kappa_i} 
\right)
\left[
1 + 
s_\alpha s_\beta 
e^{i\Phi(\omega_s^r,\omega_i^r)}
\right]
\alpha_s^\dagger(\omega_s^r) \beta_i^\dagger(\omega_i^r) \ket{0}.
\end{aligned}
\end{equation}
Here $\Phi(\omega_s^r,\omega_i^r)$ denotes the central-slot two-photon phase,
\begin{equation}
\label{equation:Central slot phase} 
\Phi(\omega_s^r,\omega_i^r)
=
\phi_s + \phi_i - \phi_p 
+ 
\sum_{j = s,i}
\omega_j^r T_j,
\end{equation} 
where $T_j$ is the overall delay mismatch for channel $j$,
\begin{equation}
\label{equation:Overall delay mismatch}
T_j
=
\tau_j - \frac{\tau_p}{\kappa_j}
=
\delta\tau_j + \Delta\tau_j^D,
\end{equation}
and where $\delta\tau_j = \tau_j - \tau_p$ is the analyzer-delay mismatch and 
\begin{equation}
\label{equation:Doppler-induced delay mismatch}
\Delta\tau_j^D
=
\left(
1 - \frac{1}{\kappa_j}
\right)
\tau_p,
\end{equation}
is the Doppler-induced delay mismatch. If the analyzer delays satisfy $\tau_j = \tau_p/\kappa_j$, Eq.~\eqref{equation:Overall delay mismatch} yields $T_j = 0$. As a result, the frequency-dependent terms in Eq.~\eqref{equation:Central slot phase} disappear, and the central-slot phase simplifies to $\phi_s + \phi_i - \phi_p$. For arbitrary $T_j$, the effect of $\Phi(\omega_s^r,\omega_i^r)$ on the analyzer outcomes is quantified by the central-slot coincidence probabilities. For each output-port pair $\alpha\beta$, the central-slot coincidence probability $\mathcal{P}_{\alpha\beta}^{r, X}$ is obtained by integrating the squared modulus of the post-selected spectral amplitude in Eq.~\eqref{equation:phase-reduced central-slot state component} over the received frequencies (see Appendix~\ref{appendix:Analyzer path expansion and central-slot projection}), 
\begin{equation} 
\begin{aligned} 
\label{equation:X-basis coincidence probability}
\mathcal{P}_{\alpha\beta}^{r, X}
&=
\frac{1}{16}
\int d\omega_s^rd\omega_i^r\frac{1}{\kappa_s\kappa_i} \Big|F^e
\Big(
\frac{\omega_s^r}{\kappa_s}, \frac{\omega_i^r}{\kappa_i} 
\Big)
\Big|^2
\Big[
1 + 
s_\alpha s_\beta 
\cos{
\big(
\Phi(\omega_s^r, \omega_i^r)
\big)}
\Big].
\end{aligned}
\end{equation}

\subsection{Doppler contribution to phase-basis correlations}
\label{subsection:Doppler contribution to phase-basis correlations}

The four central-slot coincidence probabilities given by Eq.~\eqref{equation:X-basis coincidence probability} have the same spectral weight and differ only by the sign product $s_\alpha s_\beta$ in the interference term. By grouping $CC$ and $DD$ as correlated outcomes, and $CD$ and $DC$ as anticorrelated outcomes, we define the signed phase-basis contrast as
\begin{equation}
\label{equation:Signed correlated contrast}
\mathcal{C}^X
=
\frac{
\mathcal{P}_{CC}^{r, X} 
+
\mathcal{P}_{DD}^{r, X} 
-
\mathcal{P}_{CD}^{r, X}
-
\mathcal{P}_{DC}^{r, X}
}
{
\mathcal{P}_{CC}^{r, X} 
+
\mathcal{P}_{DD}^{r, X} 
+
\mathcal{P}_{CD}^{r, X}
+
\mathcal{P}_{DC}^{r, X}
}.
\end{equation}
Substituting the output-port-pair probabilities from Eq.~\eqref{equation:X-basis coincidence probability} and noting that the Doppler transformation preserves single-pair normalization, Eq.~\eqref{equation:Signed correlated contrast} becomes (see Appendix~\ref{appendix:Gaussian phase-basis visibility})
\begin{equation}
\label{eq:Central slot signed contrast}
\mathcal{C}^X
=
\int d\omega_s^rd\omega_i^r\frac{1}{\kappa_s\kappa_i} 
\left|F^e
\left(
\frac{\omega_s^r}{\kappa_s},\frac{\omega_i^r}{\kappa_i} 
\right)
\right|^2
\cos{
\left(
\Phi(\omega_s^r,\omega_i^r)
\right)}.
\end{equation}
The phase-basis visibility is $\mathcal{V}^X = |\mathcal{C}^X|$. $\mathcal{C}^X = 1$ indicates perfectly correlated outcomes, $\mathcal{C}^X = 0$ indicates zero phase-basis contrast, and $\mathcal{C}^X = -1$ indicates perfectly anticorrelated outcomes.

Writing each received frequency as $\omega_j^r = \omega_{j0}^r + \Omega_j^r$, Eq.~\eqref{equation:Central slot phase} becomes
\begin{equation}
\label{equation:Central slot phase decomposition}
\Phi(\omega_s^r, \omega_i^r)
=
\Theta
+ 
\Omega_s^rT_s
+ 
\Omega_i^rT_i,
\end{equation} 
where 
\begin{equation}
\label{equation:Carrier phase}
\Theta
=
\phi_s + \phi_i - \phi_p 
+ 
\omega_{s0}^rT_s
+
\omega_{i0}^rT_i
\end{equation} 
represents the carrier contribution, which determines the phase-basis operating point. In contrast, as described in Sec.~\ref{subsection:Doppler transformation of the time-bin state}, the remaining detuning-dependent terms $\Omega_j^r T_j$, which include both the analyzer-delay mismatch and the Doppler-induced delay mismatch, reduce the phase-basis visibility.

\subsubsection{Zero-Doppler calibration}

Correlated phase-basis outcomes are used as a reference, defining the target zero-Doppler contrast as $\mathcal{C}^X = 1$ in the ideal case. Substituting Eq.~\eqref{equation:Overall delay mismatch} into Eq.~\eqref{equation:Carrier phase}, with $\omega^r_{j0} = \kappa_j\omega^e_{j0}$, and applying energy conservation in the emitter frame for \gls{spdc}, $\omega_{s0}^e + \omega_{i0}^e = \omega_{p0}^e$, yields
\begin{equation}
\label{equation:Carrier Phase factorized}
\Theta
= 
\left(
\kappa_s\omega_{s0}^{e}\tau_s + \phi_s
\right)
+ 
\left(
\kappa_i\omega_{i0}^{e}\tau_i + \phi_i
\right)
- 
\left(\omega_{p0}^{e}\tau_p +  \phi_p\right).
\end{equation}

Prior to the satellite pass, each \gls{umzi} is independently calibrated using a narrow-linewidth continuous-wave field at the nominal emitted carrier frequency $\omega_{j0}^e$. In the zero-Doppler configuration, $\kappa_j = 1$ and $\omega_{j0}^e = \omega_{j0}^r$. According to the \gls{umzi} transformation in Eq.~\eqref{equation:UMZI Mode Transformation}, the $C$-port intensity at the carrier frequency reaches its maximum when
\begin{equation}
\label{equation:Local Calibration Condition}
\omega_{j0}^e\tau_j + \phi_j = 0 \pmod{2\pi},\qquad j\in\{s,i\}.
\end{equation}
These local conditions determine the analyzer-phase references.

After the local phase calibration step, the zero-Doppler carrier phase satisfies $\Theta = -(\omega_{p0}^e\tau_p + \phi_p)$. To establish the ideal temporal-overlap reference, we additionally impose matched delays, $\tau_s = \tau_i = \tau_p$. At zero Doppler, this gives $T_s = T_i = 0$, so Eq.~\eqref{equation:Central slot phase decomposition} becomes $\Phi(\omega_s^r,\omega_i^r) = -(\omega_{p0}^e\tau_p + \phi_p)$. Thus, achieving $\mathcal{C}^X = 1$ additionally requires the source-phase condition
\begin{equation}
\label{equation:Pump Local Calibration Condition}
\omega_{p0}^e\tau_p + \phi_p = 0 \pmod{2\pi}.
\end{equation}

\subsubsection{Carrier-phase shift and detuning-dependent dephasing}

During the satellite pass, the pump and analyzer delays $\tau_p,\tau_j$ and calibrated phases $\phi_p,\phi_j$ remain constant and are not affected by propagation. However, the Doppler effect changes the received carrier frequencies and, consequently, the phases accumulated by the transmitted fields in the analyzers.

For matched analyzer delays, $\tau_s = \tau_i = \tau_p$, substituting the local and source-phase calibration conditions from Eqs.~\eqref{equation:Local Calibration Condition} and \eqref{equation:Pump Local Calibration Condition} into the carrier phase contribution Eq.~\eqref{equation:Carrier phase} yields
\begin{equation}
\label{equation:Carrier phase calibrated}
\Theta
=
\omega_{s0}^e(\kappa_s - 1)\tau_p
+
\omega_{i0}^e(\kappa_i - 1)\tau_p
\pmod{2\pi}.
\end{equation}
Although the local analyzer settings remain fixed, Doppler rescaling of the transmitted fields induces a shift in the carrier contribution equal to the sum of the two link-induced phases. The separable structure of Eq.~\eqref{equation:Carrier phase calibrated} allows each contribution to be estimated and compensated at its respective \gls{ogs}. When both contributions are compensated, $\Theta$ is restored to $0 \pmod{2\pi}$. The remaining central-slot phase is then entirely detuning-dependent: $\Phi = \Omega_s^r T_s + \Omega_i^r T_i$. 

Equation~\eqref{equation:Carrier phase calibrated} applies in the matched-delay configuration. For the subsequent sensitivity analysis, we examine nonzero analyzer-delay mismatches, $\delta\tau_j \neq 0$, while keeping the total carrier contribution $\Theta$ compensated. This approach isolates visibility loss due to detuning-dependent effects from carrier-phase tracking. The emitted-frame detunings satisfy $\Omega_s^e + \Omega_i^e = \Omega_p^e$. Together with $\Omega_j^r = \kappa_j\Omega_j^e$, this yields
\begin{equation}
\label{equation:Reciever detunings relation}
\Omega_i^r 
=
-
\frac{\kappa_i}{\kappa_s}\Omega_s^r
+\kappa_i\Omega_p^e.
\end{equation}
Consequently, the detuning-dependent phase can be written as
\begin{equation}
\label{equation:Spectral phase detunings}
\Phi
=
\Omega_s^rT_{\mathrm{eff}}
+\kappa_i\Omega_p^eT_i,
\end{equation}
where
\begin{equation}
\label{equation:Differential effective delay}
T_{\mathrm{eff}} 
=
T_s - \frac{\kappa_i}{\kappa_s}T_i
=
\delta\tau_s
-
\frac{\kappa_i}{\kappa_s}\delta\tau_i
+
\frac{\kappa_s-\kappa_i}{\kappa_s}\tau_p.
\end{equation}
The first term in Eq.~\eqref{equation:Spectral phase detunings} describes the phase variation along the anticorrelated signal-and-idler detuning coordinates, whereas the second describes phase variation associated with the finite pump bandwidth. Even for perfectly matched analyzer delays, $\delta\tau_s = \delta\tau_i = 0$, unequal Doppler factors thus produce a nonzero effective delay $T_{\mathrm{eff}} = \tau_p(\kappa_s - \kappa_i)/\kappa_s$, and hence a signal-detuning phase contribution.

For the Gaussian source model, the analytical expression for phase-basis visibility after carrier-phase compensation is provided in Eq.~\eqref{supplement_equation:Exact Gaussian visibility} (see Appendix~\ref{appendix:Gaussian phase-basis visibility}). The visibility depends independently on the overall delay mismatches, $T_s$ and $T_i$, attaining $\mathcal{V}^X = 1$ only when $T_s = T_i = 0$. In the narrow-pump regime, $(\sigma_{\omega,p}^I)^2 \ll (\sigma_{\omega,j}^I)^2$, the coefficient of $T_i$ in the first term approaches $\kappa_i/\kappa_s$, while the direct pump-detuning contribution becomes negligible when $\kappa_i\sigma_{\omega,p}^{I}|T_i| \ll 1$. Under these conditions, the visibility simplifies to
\begin{equation}
\label{equation:Gaussian visibility}
\mathcal{V}^X
\approx
\exp\Bigg\{
-
\frac{1}{2}
\frac{T_{\mathrm{eff}}^2}{\sigma_{\mathrm{t, eq}}^2}
\Bigg\},\quad
\frac{1}{\sigma_{\mathrm{t, eq}}^2}
=
\kappa_s^2
\frac{(\sigma_{\omega,s}^I)^2
(\sigma_{\omega,i}^I)^2
}{(\sigma_{\omega,s}^I)^2
+
(\sigma_{\omega,i}^I)^2}.
\end{equation}

Our results show that Doppler-induced carrier-phase contributions decompose into independent signal- and idler-link terms, whose temporal evolution is determined by Doppler rescaling of the received carrier frequencies. Because these terms are link-specific, we can estimate and compensate them individually at the respective \glspl{ogs}, thereby reestablishing the calibrated phase reference, $\Theta = 0 \pmod {2\pi}$. Nevertheless, this compensation does not mitigate the residual visibility loss dictated by $T_{\mathrm{eff}}$.

\subsection{BBM92 error rates}
\label{subsection:BBM92 error rates}

The phase-basis interference derived in Sec.~\ref{subsection:Doppler contribution to phase-basis correlations} is characterized by the signed contrast $\mathcal{C}^X$, whose magnitude $\mathcal{V}^X = |\mathcal{C}^X|$ quantifies the strength of the correlations, while its sign determines whether the correlations appear in the $CC$ and $DD$ output-port pairs, for $\mathcal{C}^X > 0$, or in the $CD$ and $DC$ output-port pairs, for $\mathcal{C}^X < 0$. During the satellite pass, the Doppler-induced carrier phase can reverse the sign of $\mathcal{C}^X$, thereby alternately exchanging the correlated and anticorrelated output-port parities.

In the fixed-parity approach, the zero-Doppler calibration parity is retained throughout the pass, so that correlated outcomes ($CC$, $DD$) are always regarded as correct. The intrinsic $X$-basis error probability is then $e_d^X = (1 - \mathcal{C}^X)/2$. In the relabeled-parity approach, the parity is instead updated at each sample during classical postprocessing, according to the sign of $\cos\Theta$, so that $e_d^X = (1 - \mathcal{V}^X)/2$. Because the parity depends on the combined phase $\Theta = \phi_s^D + \phi_i^D \pmod{2\pi}$ of Eq.~\eqref{equation:Carrier phase calibrated}, it cannot be determined from a local measurement at either \gls{ogs}. It can, however, be computed from the known ephemeris of both links, or obtained by exchanging over the classical channel the link phases measured locally at each \gls{ogs}. The two \glspl{ogs} must agree in advance that only one of them, here the idler \gls{ogs}, inverts its $X$-basis outcomes when $\cos\Theta < 0$. Both expressions of $e_d^X$ follow from Eq.~\eqref{equation:Signed correlated contrast}.

The time-bin $Z$-basis corresponds to coincidence detections at the first and last time slots arising from the early and late pump time bins $|ESS\rangle$ and $|LLL\rangle$, respectively. The four unequal-arm pathways are discarded during basis reconciliation, as described in Appendix~\ref{appendix:Analyzer path expansion and central-slot projection}. In the ideal temporal-filtering limit, in which detector timing jitter and the temporal overlap of adjacent pump cycles are neglected,  the two accepted time-bin outcomes are perfectly distinguishable, so the intrinsic $Z$-basis error probability is negligible, i.e., $e_d^Z = 0$.

Channel loss and background signal are included through the conditional coincidence probability following Ref.~\cite{Ma_2007}. We define the local background-detection probability at receiver $j \in \{s,\,i\}$ as $Y_{0j} = \left(N_{\mathrm{det}}D + B\right)\tau_c$, where $N_{\mathrm{det}}$ is the number of detectors, $D$ is the dark-count rate per detector, $B$ is the background-light count rate at the receiver, and $\tau_c$ is the coincidence-window duration. The coincidence detection probability, given an n-photon pair is emitted, is then
\begin{equation}
\label{eq:n-photon pair yield}
Y_n 
=
\left[1 - (1 - Y_{0s})(1 - \eta_s)^n\right]
\left[1 - (1 - Y_{0i})(1 - \eta_i)^n\right],
\end{equation}
where $\eta_j$ is the total single-photon detection efficiency of channel $j$. For the vacuum contribution, $Y_0 = Y_{0s}Y_{0i}$, whereas for a single emitted pair, $Y_1$ follows from Eq.~\eqref{eq:n-photon pair yield} for $n = 1$.

Following the detection-error model of Ma et al., the error probability for a detected coincidence in basis $b \in \{X, Z\}$, conditioned on a single-pair emission, is~\cite{Ma_2007}
\begin{equation}
\label{equation:Single-pair error}
e_1^b
=
e_0 - \frac{(e_0 - e_d^b) \eta_s \eta_i}{Y_1},
\end{equation}
where $e_0$ is the error probability associated with random background coincidences.

Let $P(n)$ denote the probability that the source generates an $n$-photon pair in one pump cycle, and let $\mu$ denote the corresponding mean photon pair number. In the low-brightness regime $\mu \ll 1$, we truncate the detection model to vacuum ($n = 0$) and single-pair contributions ($n = 1$). Therefore to first order in $\mu$, $P(0) \approx 1 - \mu$, and $P(1) \approx \mu$. The resulting \gls{qber} conditioned on basis $b \in \{X,Z\}$ is
\begin{equation}
\label{equation:Basis dependent QBER}
E^b Q 
= 
e_0 Y_0 P(0) + e_1^b Y_1 P(1),
\end{equation}
with
\begin{equation}
\label{equation:Overall coincidence gain}
Q = Y_0 P(0) + Y_1 P(1),
\end{equation}
the low-brightness form of the overall coincidence gain introduced in Ref.~\cite{Ma_2007}.

\subsection{Secret-key rate}
\label{subsection:Secret-key rate}

In the present \gls{bbm92} model, only accepted $Z$-basis (time-bin) events are used for key generation, while the accepted $X$-basis (phase) events are used to estimate the phase error that determines the privacy amplification cost. This is a modeling choice rather than a protocol requirement, since the key can also be extracted from both bases, as done in Refs.~\cite{Merolla_2023, Chapman_2022}. Within the asymptotic entanglement-based security model of Ma et al., assuming an infinitely large amount of distributed data, and thereby neglecting statistical fluctuations, the asymptotic key generation rate is bounded by~\cite{Ma_2007}
\begin{equation}
\label{equation:Secret key rate}
R \geq p_ZQ \left[1 - f_{\mathrm{EC}}H_2(E^Z) - H_2(E^X) \right],
\end{equation}
where $Q$ is the overall coincidence gain defined in Eq.~\eqref{equation:Overall coincidence gain} evaluated before temporal post-selection and basis reconciliation, $H_2(x) = -x\log_2 x - (1 - x)\log_2(1 - x)$ is the binary entropy function, and $f_{\mathrm{EC}} \geq 1$ is the bi-directional error correction efficiency ($f_{\mathrm{EC}} = 1$ corresponds to the Shannon limit). Here, the term $p_ZQ$ represents the coincidence gain associated with accepted $Z$-basis events. Following the interpretation of Pereira et al.~\cite{Pereira_2019}, the term $f_{\mathrm{EC}}H_2(E^Z)$ accounts for the information disclosed during error correction to reconcile the two end users' $Z$-basis raw keys, whereas $H_2(E^X)$ quantifies the privacy-amplification cost required to reduce the amount of information available to an eavesdropper. Consequently, the Doppler-induced degradation of the phase-basis correlations increases $E^X$, hence reducing the extractable key. With fixed parity, $E^X$ may exceed 1/2. Since $H_2$ is symmetric about 1/2, retaining these values would suggest the phase is known and that parity is relabeled whenever $E^X > 1/2$. In contrast, fixed parity assumes no phase information and holds the calibration parity throughout the pass. Therefore, we use $\min(E^X, 1/2)$ in $H_2$. In the numerical model, $p_Z$ and $p_X$, derived in Appendix~\ref{appendix:Analyzer path expansion and central-slot projection}, are evaluated directly by integrating the post-analyzer joint temporal intensity over the corresponding accepted temporal regions. 

\subsection{Finite-key length}
\label{subsection:Finite-key length}

For a finite data set acquired within a shared access time window $\Delta T$, we define $N_Z$ as the number of accepted $Z$-basis events for key generation and $N_X$ as the number of accepted $X$-basis events for phase-error estimation. The expected sample size for each basis $b \in \{Z, X\}$ is then
\begin{equation}
\label{equation:finite-key accepted events}
N_b = f_{\mathrm{rep}}\int_{\Delta T}p_b(t)Q(t)dt,
\end{equation}
where $f_{\mathrm{rep}}$ is the pump repetition rate. The corresponding expected number of erroneous detections is
\begin{equation}
\label{equation:finite-key erroneous events}
N_{\mathrm{err},b} 
=
f_{\mathrm{rep}}\int_{\Delta T}p_b(t)Q(t)E^b(t)dt.
\end{equation}
Therefore, the \gls{qber} over the acquisition interval is $\bar{E}^{b} = N_{\mathrm{err},b}/N_b$. Adapting the \gls{bbm92} finite-key analysis of Chapman et al.~\cite{Chapman_2022}, with the key extracted from the $Z$ basis only and using the correspondence $Q_{2,\mathrm{est}}^{b} = E^b$,  we evaluate the error-correction term from $\bar{E}^{Z}$, and the privacy-amplification term from $\bar{E}^{X} + \Delta(N_Z, N_X, \beta)$, where
\begin{equation}
\label{equation:finite-sample induced  statistical noise}
\Delta(N_Z,N_X,\beta)
=
\sqrt{
\frac{(N_Z + N_X)}{N_ZN_X}
\frac{(N_X + 1)}{N_X}
\ln\left(\frac{1}{\beta}\right)
},
\end{equation}
quantifies the statistical noise arising from the finite sample sizes $N_Z$ and $N_X$, and $\beta$ is an auxiliary parameter. The finite secret-key length for which the \gls{qkd} protocol is both $\varepsilon_{\mathrm{sec}}$-secret and $\varepsilon_{\mathrm{cor}}$-correct is given by~\cite{Chapman_2022},
\begin{equation}
\label{equation:secret key length}
\ell
= 
\max_{\beta \in (0, \varepsilon_{\mathrm{sec}}/4)}
\left\{
N_Z
\left[
1
-
f_{\mathrm{EC}}H_2\left(\bar{E}^Z\right)
-
H_2\left(\bar{E}^X + \Delta(N_Z, N_X, \beta)\right)
\right]
- 
\log_2\left(\frac{2}{\varepsilon_{\mathrm{cor}}}\right)
+ 
4\log_2(\beta) - 2
\right\}
\end{equation}
Here, the term $\mathrm{log}_2(2/\varepsilon_{\mathrm{cor}})$ represents the amount of information revealed during error verification, while $4\log_2(\beta) - 2$ is the finite-security hashing term. As in Eq.~\eqref{equation:Secret key rate}, we use $\min\left(\bar{E}^X + \Delta, 1/2\right)$ in $H_2$. As $\beta$ increases, $\Delta$ decreases and the hashing term $4\log_2(\beta) - 2$ increases, so the maximum in Eq.~\eqref{equation:secret key length} occurs when $\beta \to \varepsilon_{\mathrm{sec}}/4$.

\subsection{Quantum-channel efficiency}
\label{subsection:Quantum-channel efficiency}

For each quantum downlink $j \in \{s, i\}$, we decompose the time-dependent single-photon detection efficiency $\eta_j(t)$, which enters Eqs.~\eqref{eq:n-photon pair yield} and~\eqref{equation:Single-pair error}, as $\eta_j(t) = \eta_{\mathrm{geo},j}(t) \eta_{\mathrm{atm},j}(t) \eta_{\mathrm{term},j}$. Following the free-space link-budget model of Refs.~\cite{Aspelmeyer_2003, Kerstel_2018, deForgesdeParny_2023}, the geometric transmittance is expressed as
\begin{equation}
\eta_{\mathrm{geo},j}(t)
=
\frac{D_{\mathrm{R},j}^2}{d_j^2(t)
\left[
\theta_{\mathrm{diff},j}^2
+
\theta_{\mathrm{turb},j}^2(t)
\right]},
\label{eq:geometric_transmittance}
\end{equation}
with diffraction- and turbulence-induced divergence angles $\theta_{\mathrm{diff},j} = 2.44\lambda_j/D_{\mathrm{T},j}$, and $\theta_{\mathrm{turb},j}(t) = 2.1\lambda_j/r_{0,j}(t)$, respectively. $D_{\mathrm{R},j}$ denotes the receiver-telescope diameter, $D_{\mathrm{T},j}$ the transmitter-telescope diameter,  $d_j(t)$ the slant range (Fig.~\ref{fig:Dual-downlink QKD}), and $r_{0,j}(t)$ is the Fried coherence length. Consistent with the downlink approximation in Ref.~\cite{deForgesdeParny_2023}, we neglect turbulence-induced beam broadening by setting $\theta_{\mathrm{turb},j}(t) = 0$.

Atmospheric absorption and scattering are represented by the elevation-dependent transmittance $\eta_{\mathrm{atm},j}(t)= 10^{-A_{\mathrm{atm},j}^{\mathrm{dB}}(t)/10}$, with $A_{\mathrm{atm},j}^{\mathrm{dB}}(t) = A_{\mathrm{atm},0}^{\mathrm{dB}} \sec\left(\zeta_j(t)\right)$. Here, $A_{\mathrm{atm},0}^{\mathrm{dB}}$ represents the atmospheric attenuation at zenith, and $\zeta_j(t) = \pi/2 -\varepsilon_j(t)$ denotes the zenith angle (Fig.~\ref{fig:Dual-downlink QKD}).

The term $\eta_{\mathrm{term},j} = T_{\mathrm{T},j} T_{\mathrm{R},j} \left(1 - L_{\mathrm{P},j}\right) T_{\mathrm{opt},j} \mathrm{PDE}_j$ describes the overall efficiency of the optical receiver and detection system for link $j$. Here, $T_{\mathrm{T},j}$ and $T_{\mathrm{R},j}$ are the transmitter and receiver telescope transmissions, $L_{\mathrm{P},j}$ is the pointing loss, $T_{\mathrm{opt},j}$ is the remaining optical transmission, and $\mathrm{PDE}_j$ is the photon-detection efficiency (we presume the use of superconducting nanowire single-photon detectors (SNSPDs)). These quantities are assumed to be constant during the simulated pass.

The numerical values in Table~\ref{tab:channel_params} apply to both downlinks, so that $\eta_s(t)$ and $\eta_i(t)$ differ only in their wavelengths and in their slant ranges and elevation angles. The latter are obtained from the orbital ephemerides of the representative satellite dual-downlink scenario defined in the next section.

\begin{table}[htbp]
\centering
\caption{Quantum-link parameters used in the numerical analysis.}
\label{tab:channel_params}
\begin{tabular}{lrl}
\hline
Symbol 
& Value 
& Description \\
\hline
$D_{\mathrm{T}}$ & 0.30~m & Transmitter-telescope diameter \\
$D_{\mathrm{R}}$ & 0.80~m & Receiver-telescope diameter \\
$A_{\mathrm{atm,0}}^{\mathrm{dB}}$ & 2~dB & Zenith atmospheric attenuation \\
$T_{\mathrm{T}}$ & 0.80 & Transmitter-telescope transmission \\
$T_{\mathrm{R}}$ & 0.80 & Receiver-telescope transmission \\
$L_{\mathrm{P}}$ & 0.20 & Pointing-loss fraction \\
$T_{\mathrm{opt}}$ & 0.35 & Optical-module transmission \\
$\mathrm{PDE}$ & 0.90 & Photon-detection efficiency\\
\hline
\end{tabular}
\end{table}

\section{Orbit, ground stations, and common access window}
\label{section:Orbit, ground stations, and common access window}

In this section, we evaluate the link transmittance and the Doppler effects on dual-downlink time-bin \gls{bbm92} \gls{qkd} using a sun-synchronous orbit at an altitude of 602~km propagated over seven days. Signal and idler photons are sent to Grasse and Paris, respectively. We assign $j = s$ to the Grasse downlink and $j = i$ to the Paris downlink. Table~\ref{tab:Orbital Parameters} lists the orbital parameters, and Table~\ref{tab:ogs_coordinates} gives the \gls{ogs} coordinates. This geometry is identical to that used in our previous simulations \cite{deForgesdeParny_2023}, except that the \gls{ogs} previously located in Nice has been moved to Grasse, where the Optical Metrology laser station (MéO) is currently situated.

\begin{table}[htbp]
\centering
\caption{SIMU-CIC orbital parameters.}
\begin{tabular}{ll}
\hline
Parameter & Value \\
\hline
Simulation start (UTC) & 2021-01-01 00:00:00 \\
Simulation end (UTC) & 2021-01-08 00:00:00 \\
Altitude (km) & 602 \\
Eccentricity & $1.23\times10^{-3}$ \\
Inclination (deg) & 97.795 \\
Argument of perigee (deg) & 154.940 \\
Argument of latitude (deg) & 32.307 \\
Mean local time of ascending node (h) & 1.000 \\
Mean $J_2$ orbital period (h) & 1.613 \\
Type & Sun-synchronous orbit (SSO) \\
Model & Secular-$J_2$ \\
\hline
\end{tabular}
\label{tab:Orbital Parameters}
\end{table}

\begin{table}[htbp]
\centering
\caption{\gls{ogs} geodetic coordinates.}
\begin{tabular}{lrrr}
\hline
Station
& Longitude ($^\circ$) 
& Latitude ($^\circ$) 
& Altitude (m) \\
\hline
Paris  & 2.3510  & 48.8560  & 30 \\
Grasse & 6.9216  & 43.7546  & 1323 \\
\hline
\end{tabular}
\label{tab:ogs_coordinates}
\end{table}

Orbital ephemerides are generated with the orbital simulation package developed at the French Space Agency (CNES), known as SIMU-CIC (version 1.2.1~\cite{SIMUCIC_UserManual}), running in the Scilab (version 6.1.1) numerical computation environment. Using a 1~s time step, the simulation outputs at each sampling instant $t_k$ the elevation angle $\varepsilon_j(t_k)$, slant range $d_j(t_k)$, and slant-range rate $\dot{d}_j(t_k)$, for both downlinks. We perform the numerical analysis on the simulation ephemeris grid without interpolating between successive samples. The slant-range rate determines the Doppler stretching factor $\kappa_j(t_k)$ as defined in Eq.~\eqref{equation:Doppler Factor}. The values $d_j(t_k)$ and $\varepsilon_j(t_k)$ are used in the channel transmittance model described in Sec.~\ref{subsection:Quantum-channel efficiency}. We retain only samples for which both downlinks $j\in\{s,i\}$ simultaneously satisfy $\varepsilon_j(t_k) \geq 20^\circ$. This common-access criterion excludes low-elevation portions of the pass, where increased slant range and atmospheric path length increase link attenuation. Over the seven-day simulation, this criterion yields 18 common-access windows, for a total acquisition time of about $61~\mathrm{min}$, with individual windows lasting between $28$ and $270~\mathrm{s}$. We evaluate the combined carrier-phase evolution for each window. Among them, orbit no.~9 is the most demanding pass, with the largest maximum phase rate $|\dot{\Theta}(t)|$ and the largest total phase excursion. We therefore select it for the detailed analysis below. For this pass, the simultaneous elevation constraint gives $\Delta T = 270~\mathrm{s}$.

\begin{figure}[htbp]
\centering
\includegraphics[width=.7\linewidth]{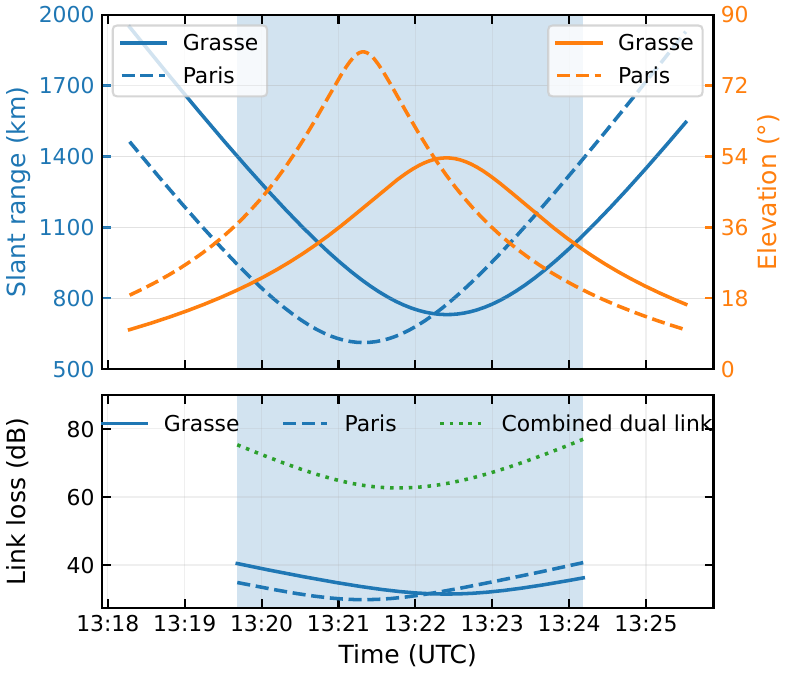}
\caption{Elevation angle $\varepsilon_j(t)$ and slant range $d_j(t)$ (\textbf{top}) and total link loss $-10\log_{10}\eta_j(t)$ together with the combined dual link (\textbf{bottom}), for the Grasse (signal) and Paris (idler) downlinks during the selected pass (orbit no.~9). The shaded region denotes the 270-s common-access window within which both downlinks satisfy $\varepsilon_j(t) \geq 20^\circ$. }
\label{fig:Valid Elevation And Slant Range}
\end{figure}

Figure~\ref{fig:Valid Elevation And Slant Range} presents the elevation angle and slant range for the selected pass, along with the link loss, which ranges from approximately $30$ to $41~\mathrm{dB}$ per downlink. The terminal efficiency $\eta_{\mathrm{term}}$ contributes a constant $7.9~\mathrm{dB}$ to this loss. The two links differ because of their distinct viewing geometries.

\section{Doppler-induced performance and phase-correction requirements}
\label{section:Doppler-induced performance and phase-correction requirements}

\subsection{Doppler-factor and carrier-phase evolution}
\label{subsection:Doppler-factor and carrier-phase evolution}

\begin{figure}[htbp]
\centering\includegraphics[width=.49\linewidth]{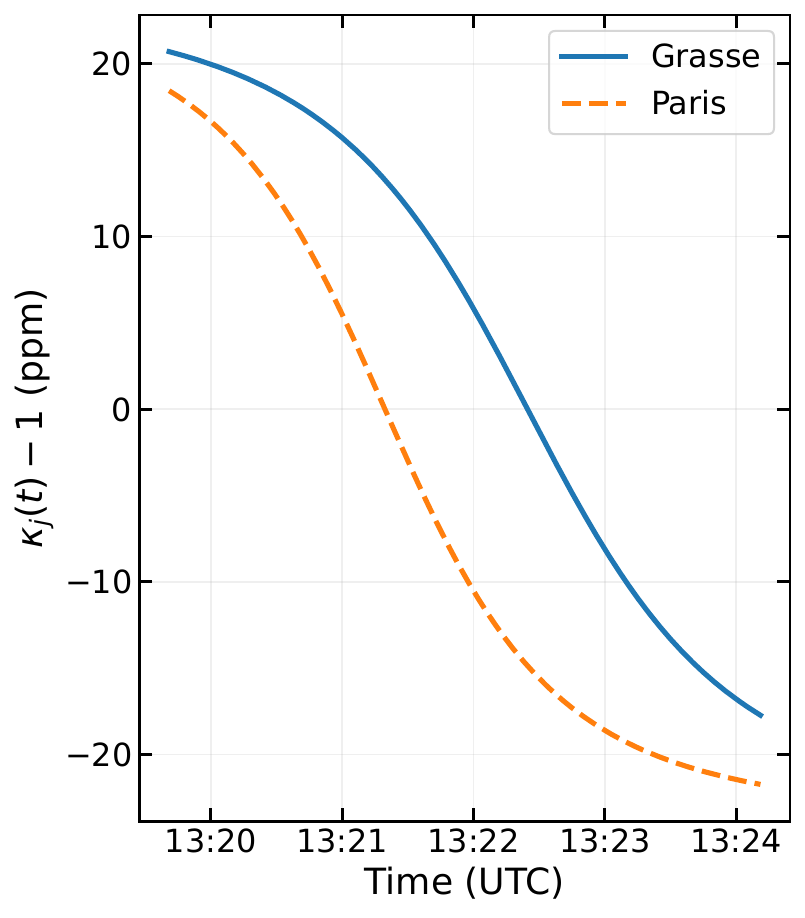}
\includegraphics[width=.49\linewidth]{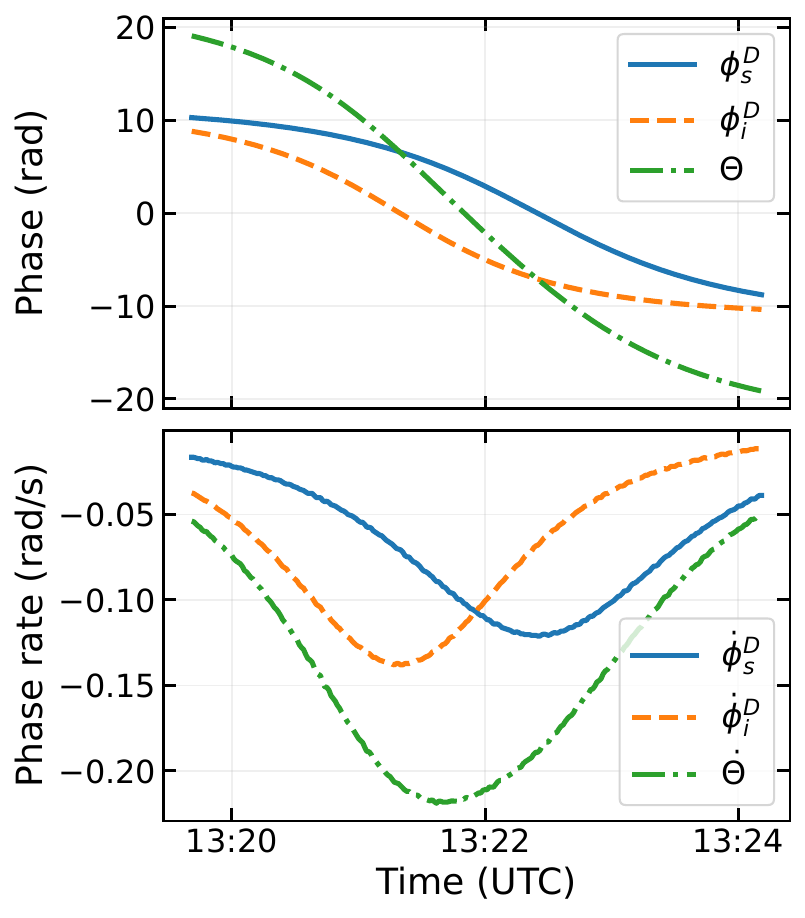}
\caption{\textbf{Left -} Doppler stretching $\kappa(t) - 1$ for the Grasse (signal) and Paris (idler) downlinks over the common-access window. Positive and negative values correspond to satellite approach and recession, respectively. \textbf{Top right -} Doppler-induced carrier phases $\phi_s^D(t)$ and $\phi_i^D(t)$ for the Grasse (signal) and Paris (idler) downlinks, respectively, together with their sum $\Theta(t) = \phi_s^D(t) + \phi_i^D(t)$ over the common-access window. \textbf{Bottom right -} Corresponding phase rates $\dot{\phi}_s^D(t)$, $\dot{\phi}_i^D(t)$, and $\dot{\Theta}(t)$. The maximum single-link and combined phase rates are approximately $\max_{j,\, t}|\dot{\phi}_j^D(t)| \simeq 0.14~\mathrm{rad \cdot s^{-1}}$ and $\max_t|\dot{\Theta}(t)| \simeq 0.22~\mathrm{rad \cdot s^{-1}}$, respectively.}
\label{fig:Doppler evolution}
\end{figure}

The Doppler-induced stretching offset, $\kappa_j(t) - 1$, is computed from the SIMU-CIC ephemerides for both downlinks. Fig.~\ref{fig:Doppler evolution} illustrates its evolution, along with the link-dependent phase $\phi_j^D(t)$ and phase rate $\dot\phi_j^D(t)$, over the common-access window for the selected orbit. The differences in their temporal evolution between the Grasse and Paris \gls{ogs} result from the distinct viewing geometries. According to the sign convention in Eq.~\eqref{equation:Doppler Factor}, $\kappa_j(t) - 1$ is positive during approach and negative during recession. Throughout the common-access window, $|\kappa_j(t) - 1| < 22~\mathrm{ppm}$. For $\tau_p = 400~\mathrm{ps}$ as specified in Table~\ref{tab:source_params}, the corresponding Doppler-induced delay mismatch defined in Eq.~\eqref{equation:Doppler-induced delay mismatch} is bounded by $\left| \Delta\tau_j^D(t) \right| \lesssim  8.8~\mathrm{fs}$. Therefore, Doppler rescaling produces a negligible change in the time-bin separation. However, although exceedingly small, multiplication of $\left| \Delta\tau_j^D(t) \right|$ by the received optical carrier frequency $\omega_{j0}^r = \kappa_j\omega_{j0}^e$ produces a significant phase shift. Evaluating Eq.~\eqref{equation:Doppler phase emitter frame} at the carrier frequency yields
\begin{equation}
\label{equation:Doppler phase emitter frame carrier}
\phi_j^D(t)
= 
\omega_{j0}^r\,\Delta\tau_j^D(t)
=
\omega_{j0}^e
\left(\kappa_j(t) - 1\right)
\tau_p,
\end{equation}
and for the signal and idler wavelengths considered here, Fig.~\ref{fig:Doppler evolution} shows that $\left|\phi_j^D(t)\right| < 11~\mathrm{rad}$.

We now verify the quasi-static approximation of Sec.~\ref{subsection:Doppler transformation of the time-bin state}, namely the assumption that $\kappa_j(t)$ can be treated as constant over the time-bin separation $\tau_p$. Differentiating Eq.~\eqref{equation:Doppler phase emitter frame carrier} with respect to the pass time $t$ yields
\begin{equation}
\label{equation:Doppler phase rate}
\dot{\phi}_j^D(t)
=
\omega_{j0}^e\tau_p\dot{\kappa}_j(t).
\end{equation}
The maximum single-link phase rate over the common-access window is $\max_{j,t} \left| \dot{\phi}_j^D(t) \right| \simeq 0.14~\mathrm{rad \cdot s^{-1}}$. Using Eq.~\eqref{equation:Doppler phase rate}, the change in Doppler factor over $\tau_p = 400~\mathrm{ps}$ is bounded by $\left|\Delta\kappa_j\right| \lesssim \max_{j,t}\left|\dot{\kappa}_j(t)\right|\tau_p \simeq 1.2\times10^{-16}$, which is negligible compared to $|\kappa_j(t) - 1| < 22\times10^{-6}$. Evaluating the early-late carrier phase with a single Doppler factor therefore introduces an error of at most $\left|\Delta\phi_j^D\right| \lesssim \max_{j,t}|\dot{\phi}_j^D(t)|\,\tau_p \simeq 5.5\times10^{-11}~\mathrm{rad}$, which is negligible compared with the $\sim 11$-rad Doppler phase. Thus, although the line-of-sight velocity produces a large variation in carrier phase, the line-of-sight acceleration leaves it essentially unchanged over $\tau_p$, which confirms the quasi-static approximation.

\begin{figure}[htbp]
\centering
\includegraphics[width=.7\linewidth]{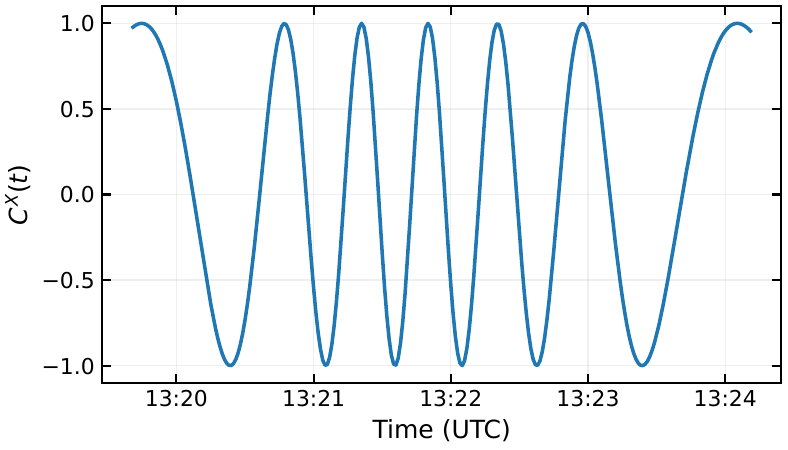}
\caption{Signed phase-basis contrast $\mathcal{C}^X(t)$ over the common-access window. Positive and negative values correspond to correlated and anticorrelated output-port parity, respectively.}
\label{fig:Signed Contrast Uncorrected}
\end{figure}

With matched delays and zero-Doppler calibration, Eq.~\eqref{equation:Carrier phase calibrated} defines the central-slot carrier contribution $\Theta(t)$ as the sum of the two channel-induced Doppler phases. Figure~\ref{fig:Doppler evolution} shows that during the common-access window $\Theta(t)$ undergoes a total phase excursion of approximately 38~rad, corresponding to about six $2\pi$ cycles. Fig.~\ref{fig:Signed Contrast Uncorrected} shows the resulting uncorrected phase-basis response, where the signed contrast $\mathcal{C}^X(t)$ oscillates between correlated and anticorrelated outcomes as the carrier phase sweeps across successive interference fringes. 

These results show that Doppler effects have minimal influence on time-bin separation but strongly modify carrier-phase evolution and consequently the $X$-basis interference. 

\subsection{Sensitivity to analyzer-delay mismatch}
\label{subsection:Sensitivity to analyzer-delay mismatch}

Equation~\eqref{equation:Gaussian visibility} establishes that spectral dephasing is governed by $|T_{\mathrm{eff}}|/\sigma_{t,\mathrm{eq}}$, with $T_{\mathrm{eff}}$ including both analyzer-delay mismatch and differential Doppler stretching. For the source parameters of Table~\ref{tab:source_params}, $\sigma_{t,\mathrm{eq}} \simeq 4.25~\mathrm{ps}$. In the zero-Doppler limit ($\kappa_s = \kappa_i = 1$), Eq.~\eqref{equation:Differential effective delay} reduces to $T_{\mathrm{eff}} = \delta\tau_s - \delta\tau_i$. A differential delay mismatch of 1\% relative to the nominal $\tau_p = 400~\mathrm{ps}$ yields $|T_{\mathrm{eff}}| = 4~\mathrm{ps}$ and $|T_{\mathrm{eff}}|/\sigma_{t,\mathrm{eq}} \simeq 0.94$. According to Eq.~\eqref{equation:Gaussian visibility}, this corresponds to a phase-basis visibility of $\mathcal{V}^X \simeq 0.64$. For this mismatch, the associated spectral phase slope at $\lambda_s = 1520~\mathrm{nm}$ is
\begin{equation}
\left|
\frac{\partial\Phi_{\mathrm{UMZI}}}{\partial\lambda}
\right|
=
\frac{2\pi c}{\lambda_s^2}|T_{\mathrm{eff}}|
\simeq
3.26~\mathrm{rad \cdot nm^{-1}}.
\end{equation}
Thus, even a percent-level differential analyzer-delay mismatch causes significant visibility loss and a $\pi$-scale phase variation over a wavelength interval of about 1~nm. The differential delay must therefore be controlled. A target visibility $\mathcal{V}^X \geq 0.98$ requires the \gls{umzi} delays at the two \glspl{ogs} to match within 0.85~ps, which defines the engineering tolerance of the analyzers.

Having established this sensitivity to analyzer-delay mismatch, we now isolate the contribution of differential Doppler stretching. With matched analyzers ($\delta\tau_s = \delta\tau_i = 0$), the effective mismatch of Eq.~\eqref{equation:Differential effective delay} reduces to $T_{\mathrm{eff}} = \Delta\tau_s^D - \frac{\kappa_i}{\kappa_s}\Delta\tau_i^D$. Therefore, only the difference in Doppler stretching between the two downlinks contributes to the residual spectral dephasing. Since $|\Delta\tau_j^D| \lesssim 8.8~\mathrm{fs}$ over the common-access window, $|T_{\mathrm{eff}} |\lesssim 17.6~\mathrm{fs}$. The resulting spectral phase slope at $\lambda_s = 1520~\mathrm{nm}$ is bounded by
\begin{equation}
\left|
\frac{\partial\Phi_D}{\partial\lambda}
\right|
=
\frac{2\pi c}{\lambda_s^2}|T_{\mathrm{eff}}|
\lesssim
1.4\times10^{-2}~\mathrm{rad \cdot nm^{-1}}.
\end{equation}
Thus, once the analyzer delays are sufficiently matched, the Doppler-induced detuning-dependent contribution becomes negligible. The carrier Doppler phase then remains the primary correction target.

\subsection{Carrier-phase correction-rate requirement}
\label{subsection:Carrier-phase correction-rate requirement}
 
Assuming matched analyzer delays, $\tau_s = \tau_i = \tau_p$, and neglecting the residual detuning-dependent Doppler-induced dephasing quantified above, we now determine the rate required to track the time-dependent carrier phase for the selected pass (orbit no.~9). We model the correction on each downlink as an ideal sample-and-hold operation. At each update time $t_k$, a time-bin-selective phase modulator at each \gls{ogs} applies a phase $\phi_j^{\mathrm{PM}}$ to the late time bin. The received state of
Eq.~\eqref{equation:Received Time-Bin State} then becomes 
\begin{equation}
\label{equation:Received Time-Bin State with PM}
\begin{aligned}
\ket{\Psi^r}
&=
\frac{1}{\sqrt{2}}
\int d\omega_s^r d\omega_i^r
\frac{1}{\sqrt{\kappa_s\kappa_i}}
F^e
\left(
\frac{\omega_s^r}{\kappa_s},
\frac{\omega_i^r}{\kappa_i}
\right)\\
&\quad\times
\left[
1 +
\exp\left\{
i(\omega_s^r + \omega_i^r)\tau_p
+
i\phi_p
-
i\sum_{j = s,i}\left(\phi_j^D(\omega_j^r)
- \phi_j^{\mathrm{PM}}\right)
\right\}
\right]
a_s^\dagger(\omega_s^r)
a_i^\dagger(\omega_i^r)
\ket{0}.
\end{aligned}
\end{equation}
Compensating for the link-dependent instantaneous Doppler-induced carrier phase of Eq.~\eqref{equation:Doppler phase emitter frame carrier} therefore requires $\phi_j^{\mathrm{PM}}(t_k) = \phi_j^D(t_k)$, the detuning-dependent term of Eq.~\eqref{equation:Doppler phase receiver frame} being negligible for matched delays (Sec.~\ref{subsection:Sensitivity to analyzer-delay mismatch}). This command is held fixed until $t_{k + 1} = t_k + T_{\mathrm{corr}}$, with $T_{\mathrm{corr}} = 1/f_{\mathrm{corr}}$. We assume ideal phase correction and negligible actuation latency, so that system requirements are governed exclusively by the phase-evolution rate, irrespective of the phase estimation protocol. We further assume that both \glspl{ogs} update their corrections at the same times $t_k$, so that  $\phi_j^{\mathrm{PM}}(t) = \phi_j^D(t_k)$ for
$t_k \leq t < t_{k + 1}$. Given Eq.~\eqref{equation:Received Time-Bin State with PM}, the combined residual carrier phase $\Delta\Theta_k(t) = \sum_{j}\left[\phi_j^D(t) - \phi_j^{\mathrm{PM}}(t)\right]$ becomes
\begin{equation}
\label{equation:combined-sample-and-hold-residual}
\Delta\Theta_k(t)
=
\Theta(t) - \Theta(t_k)
=
\left[\phi_s^D(t) - \phi_s^D(t_k)\right]
+
\left[\phi_i^D(t) - \phi_i^D(t_k)\right],
\qquad
t_k\leq t<t_{k + 1}.
\end{equation}
Then, by the mean-value theorem there exists $\xi_k \in (t_k, t)$ such that $\Delta\Theta_k(t) = \dot{\Theta}(\xi_k) (t - t_k)$. Since $t - t_k \leq T_{\mathrm{corr}}$, the combined residual is bounded by
\begin{equation}
\label{equation:combined-residual-phase-bound}
\left|\Delta\Theta_k(t)\right|
\leq
\max_t\left|\dot{\Theta}(t)\right| T_{\mathrm{corr}},
\qquad
t_k\leq t<t_{k+1}.
\end{equation}
Given Eq.~\eqref{eq:Central slot signed contrast} and that $\mathcal{V}_k^X(t) = |\mathcal{C}_k^X(t)|$, the visibility can be written
\begin{equation}
\label{equation:residual-visibility}
\mathcal{V}_k^X(t)
\simeq
\left|\cos\left(\Delta\Theta_k(t)\right)\right|,
\qquad
t_k\leq t<t_{k+1}.
\end{equation}
For a given desired minimum phase-basis visibility $\mathcal{V}_k^X$, the maximum allowable magnitude of the combined residual phase is $\Delta\Theta_{\max} = \arccos\left(\mathcal{V}_k^X\right)$. Combining this condition with Eq.~\eqref{equation:combined-residual-phase-bound} gives the corresponding maximum update period, $T_{\mathrm{corr}}^{\max} = \arccos\left(\mathcal{V}_k^X\right)/\max_t|\dot{\Theta}(t)|$, and minimum update rate, $f_{\mathrm{corr}}^{\min} = \max_t|\dot{\Theta}(t)|/\arccos\left(\mathcal{V}_k^X\right)$. For orbit no.~9, the phase-rate evolution shown in Fig.~\ref{fig:Doppler evolution} reaches $\max_t|\dot{\Theta}(t)| \simeq 0.22~\mathrm{rad \cdot s^{-1}}$. Table~\ref{tab:visibility-correction-rate} translates this maximum phase-evolution rate into update-rate requirements for several representative phase-basis visibility targets. 

\begin{table}[htbp]
\centering
\caption{\label{tab:visibility-correction-rate} Maximum allowable combined residual phase and corresponding minimum sample-and-hold update requirements evaluated for orbit no.~9.}
\begin{tabular}{ccc}
\hline
$\mathcal{V}^X$
&
$\Delta\Theta_{\max}$ (rad)
&
$f_{\mathrm{corr}}^{\min}$ (Hz)\\
\hline
0.99 & 0.14 & 1.55 \\
0.98 & 0.20 & 1.10 \\
0.95 & 0.32 & 0.69 \\
\hline
\end{tabular}
\end{table}

For the 1~Hz update rate used in the numerical analysis below, consistent with the 1~s ephemeris sampling, this corresponds to a minimum phase-basis visibility of $\mathcal{V}_{\min}^X \simeq 0.976$.

\section{Ephemeris-assisted witness-field phase correction}
\label{section:Ephemeris-assisted witness-field phase correction}

\begin{figure}[htbp]
    \centering
    \includegraphics[width=.7\linewidth]{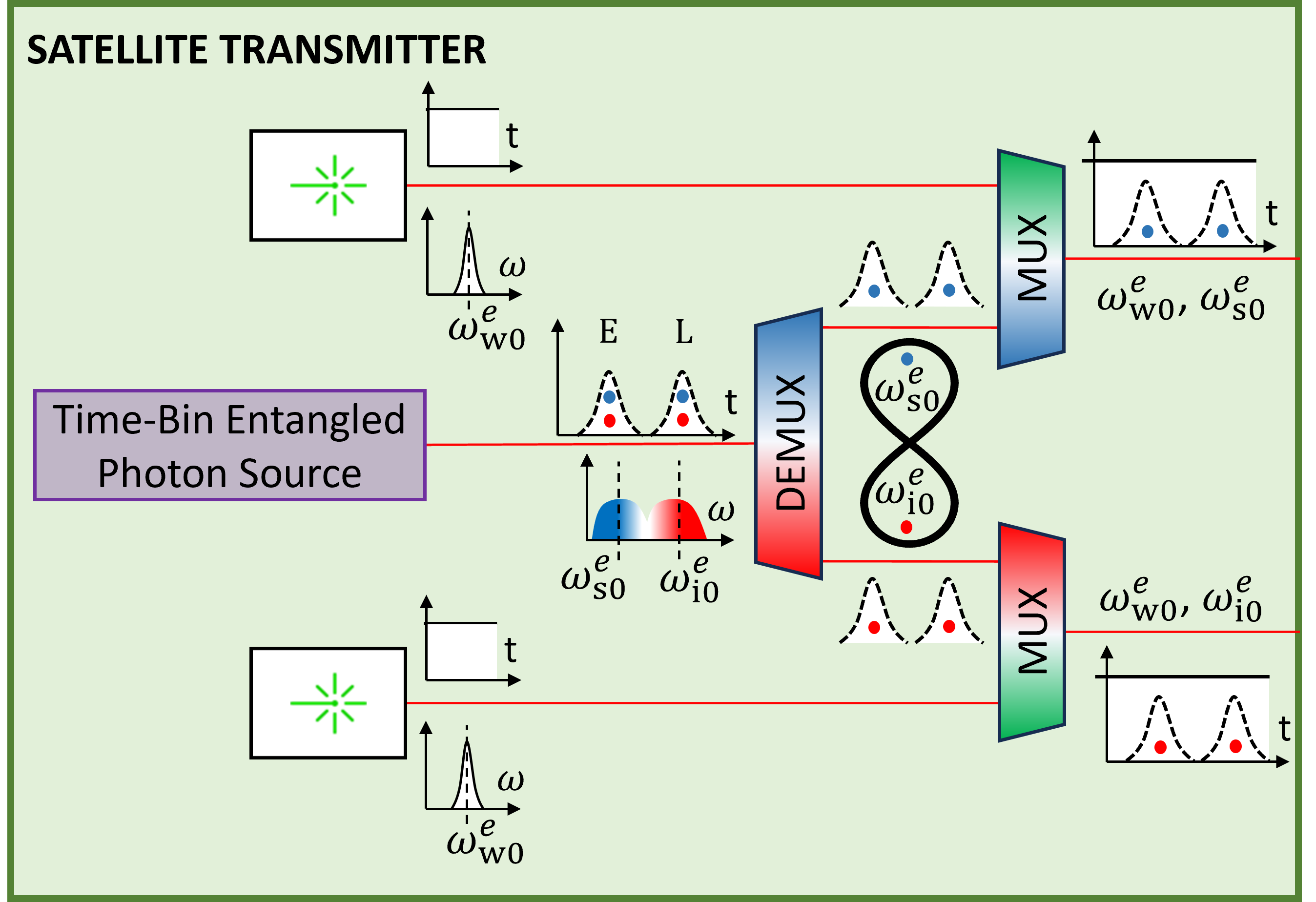}\\
\includegraphics[width=.7\linewidth]{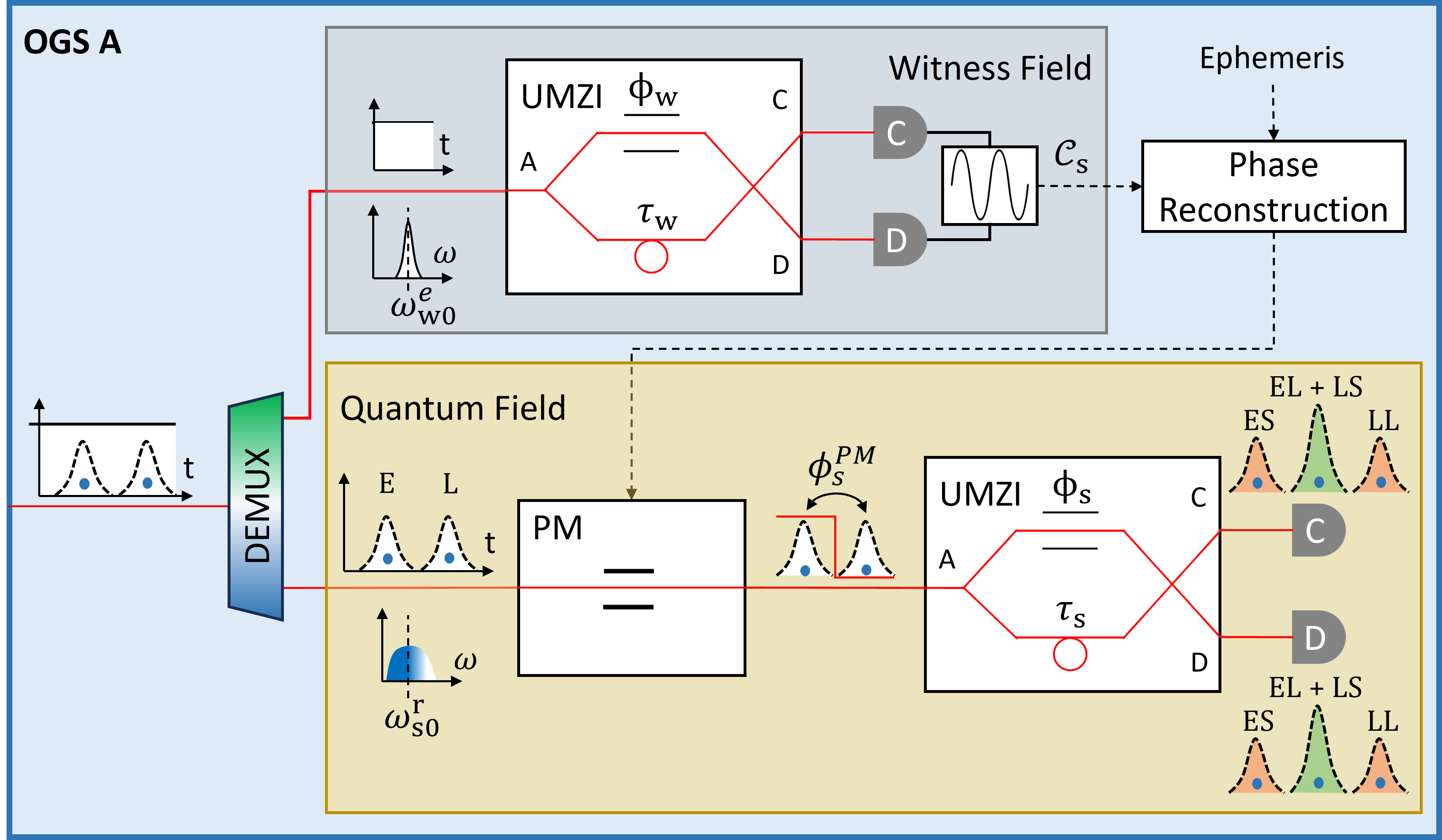}
    \caption{Proposed source and receiver architecture for local Doppler-phase correction. (\textbf{Top}) Satellite transmitter: after wavelength demultiplexing, each quantum channel is combined with a narrow-linewidth classical witness field at carrier frequency $\omega_{w0}^{e}$ using a wavelength multiplexer. The witness and quantum fields then co-propagate along the corresponding signal and idler downlinks and therefore experience the same link-dependent Doppler factor. (\textbf{Bottom}) Signal receiver at \gls{ogs} A: after wavelength demultiplexing, the received classical witness field is analyzed by an auxiliary \gls{umzi} with delay $\tau_w$ and phase $\phi_w$. The two photodetector outputs provide the normalized contrast $\mathcal{C}_s$, which is combined with the orbital ephemeris to reconstruct the local Doppler-induced witness phase. The corresponding quantum-channel phase correction is then applied by a time-bin-selective phase modulator with command $\phi_s^{\mathrm{PM}}$ acting on the late quantum time bin, before the signal photon enters the \gls{bbm92} analyzer \gls{umzi} of delay $\tau_s$ and phase $\phi_s$. The analyzer outputs $C$ and $D$ resolve the outer $ES$ and $LL$ time slots associated with the $Z$ basis and the central $EL+LS$ interference slot associated with the $X$ basis. The idler receiver has the same architecture with $s \rightarrow i$.}
    \label{fig:Witness architecture}
\end{figure}

Section~\ref{subsection:Carrier-phase correction-rate requirement} assumed that the Doppler-induced phase is known at each update. This phase can be predicted from the ephemeris or, as proposed here, measured optically at each receiver. Figure~\ref{fig:Witness architecture} illustrates the proposed architecture. Because each classical witness field co-propagates with its corresponding quantum signal, it experiences the same Doppler factor $\kappa_j(t)$. With the auxiliary \gls{umzi} set at quadrature before the pass and then held fixed, the normalized witness contrast is
$\mathcal{C}_j(t) \simeq -\sin\left(\phi_{w,j}^D(t)\right)$
(Appendix~\ref{appendix:Witness-field phase reconstruction}), where
\begin{equation}
\phi_{w,j}^D(t) = \omega_{w0}^e \left(\kappa_j(t) - 1\right)\tau_w,
\label{equation:witness phase}
\end{equation}
is the Doppler-induced witness phase. Since $\sin\phi = \sin(\pi - \phi)$, a single contrast measurement does not uniquely determine $\phi_{w,j}^D$. This branch ambiguity is resolved at each reconstruction step from the sign of the metric $\mathcal{B}_j(t_k) = \dot{\mathcal{C}}_j(t_k)\dot{\kappa}_j(t_k)$, which combines the measured contrast evolution with the ephemeris-derived Doppler factor. The selected phase is then temporally unwrapped to recover the continuous phase change, except for an unknown constant integer multiple of $2\pi$, the absolute fringe order. We determine this fringe order once by comparing the initial reconstructed phase with the ephemeris-predicted phase and retain this value throughout the uninterrupted reconstruction interval. The ephemeris is thus used only for branch selection and this one-time initialization, whereas the subsequent time-dependent phase is reconstructed from the measured witness contrast. The complete procedure is given in Appendix~\ref{appendix:Witness-field phase reconstruction}.

\begin{figure}[htbp]
    \centering
    \includegraphics[width=.7\columnwidth]{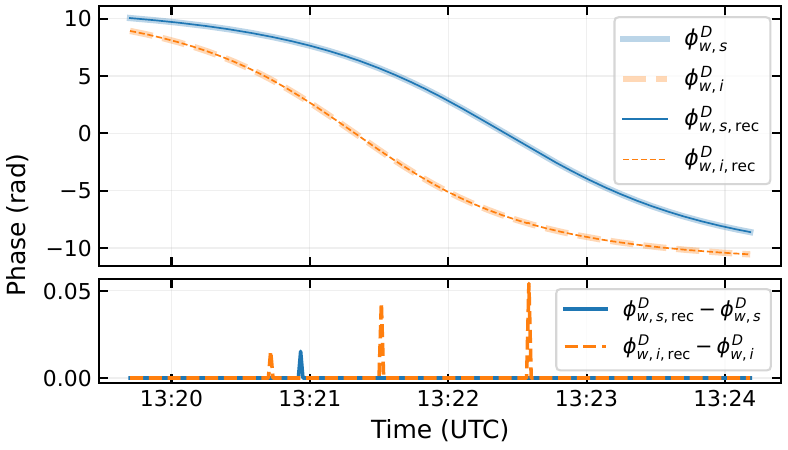}
    \caption{Ephemeris-assisted reconstruction of the Doppler-induced witness phase.
    \textbf{Top:} reconstructed phase $\phi_{w,j,\mathrm{rec}}^D$ and ephemeris-predicted phase $\phi_{w,j}^D$ for the signal ($j = s$) and idler ($j = i$) downlinks. \textbf{Bottom:} reconstruction residual $\phi_{w,j,\mathrm{rec}}^D - \phi_{w,j}^D$ for both downlinks. In the noise-free model, the residual vanishes except at four isolated samples where the branch selection is delayed by one update near a contrast extremum.}
    \label{fig:witness-phase-reconstruction}
\end{figure}

Figure~\ref{fig:witness-phase-reconstruction} shows the resulting phase reconstruction for $\lambda_{w0} = 1550~\mathrm{nm}$ and $\tau_w = 400~\mathrm{ps}$. Since the model excludes measurement noise, ephemeris error, and laser and interferometer drift, the reconstructed phase $\phi^D_{w,j,\mathrm{rec}}$ coincides with the ephemeris-predicted phase $\phi_{w,j}^D$ throughout the common-access window, which confirms the internal consistency of the reconstruction procedure. The only exceptions are four isolated samples, one on the signal link and three on the idler link, where the metric $\mathcal{B}_j$ selects the other branch near a contrast extremum, producing reconstruction errors of at most $0.05~\mathrm{rad}$. Consequently, up to these transients, the optical-correction results in Sec.~\ref{section:QKD performance with and without carrier-phase correction} apply equally to a feed-forward command derived from the ephemeris alone.

Since the witness and quantum-channel phases share the same Doppler factor, Eqs.~\eqref{equation:Doppler phase emitter frame carrier} and~\eqref{equation:witness phase} differ only through their carrier frequencies and through the witness interferometer delay $\tau_w$, which does not necessarily equal $\tau_p$. The reconstructed witness phase thus provides the local time-bin-selective phase-modulator command
\begin{equation}
\label{equation:main-witness-PM}
\phi_j^{\mathrm{PM}}(t_k) =  \frac{\omega_{j0}^e\tau_p}{\omega_{w0}^e\tau_w}\phi_{w,j,\mathrm{rec}}^D(t_k),
\end{equation}
which equals $\phi_j^D(t_k)$ when the branch and fringe order are correctly identified. Because the branch selection uses two consecutive witness samples, witness-derived updates start at $t_1$, and the command at $t_0$ is set from the ephemeris-predicted quantum-channel phase $\phi_j^{\mathrm{PM}}(t_0) = \omega_{j0}^e\left(\kappa_j(t_0) - 1\right)\tau_p$. The command is applied using the sample-and-hold protocol of Sec.~\ref{subsection:Carrier-phase correction-rate requirement}, and the two local corrections together cancel the carrier phase of Eq.~\eqref{equation:Carrier phase calibrated}.

\section{QKD performance with and without carrier-phase correction}
\label{section:QKD performance with and without carrier-phase correction}

\subsection{Single-pass QKD performance}

\begin{figure}[htbp]
\centering
\includegraphics[width=.7\columnwidth]{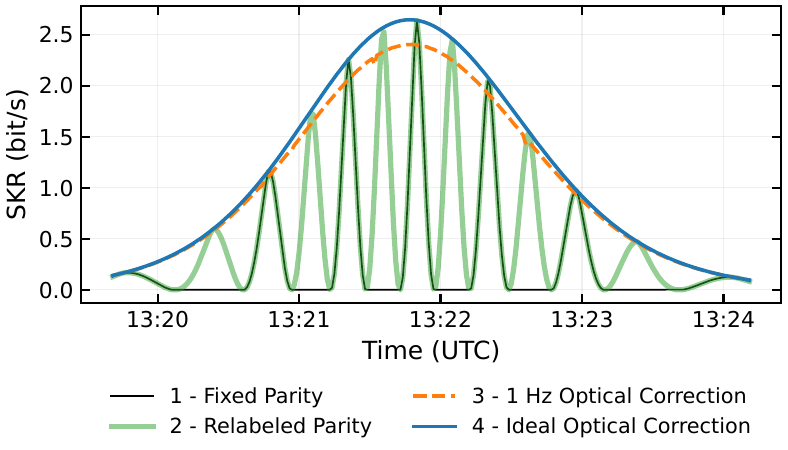}
\caption{Asymptotic secret-key rate over the common-access window of orbit no.~9 for the four treatments: (1) fixed parity and (2) relabeled parity, using the combined phase $\Theta$ computed from the ephemeris, both without optical correction, and (3) 1~Hz and (4) ideal optical correction. The ideal correction defines the loss-dependent rate envelope, which is closely followed by the 1~Hz sample-and-hold correction, whereas the rates without optical correction periodically vanish, and the fixed-parity rate retains only every other peak of the relabeled-parity rate.}
\label{fig:secret-key-rate}
\end{figure}

We compare four treatments of the Doppler-induced phase over the common-access window: (1) fixed parity, (2) relabeled parity, (3) sample-and-hold correction operating at $f_{\mathrm{corr}} = 1~\mathrm{Hz}$, and (4) ideal instantaneous correction in which the Doppler-induced phase is known exactly and compensated continuously ($T_{\mathrm{corr}} \to 0$), so that $\Theta = 0 \pmod{2\pi}$ at every instant. All other parameters, including the orbital geometry, source parameters, link efficiencies, and detection-error parameters, are identical across the four treatments, and are listed in Table~\ref{tab:qkd-model-parameters}. Figure~\ref{fig:secret-key-rate} presents the computed asymptotic secret-key rates, and Table~\ref{tab:qkd-performance-summary} provides the pass-averaged $X$-basis \gls{qber} and integrated asymptotic secret-key yield for orbit no.~9. The background rate $B$ is assumed constant and is representative of nighttime operation, or daytime operation with very tight spectral filtering. Orbit no.~9 occurs during daylight, when sky background counts can be expected to be higher. However, this pass was selected to assess the most challenging Doppler dynamics, not to represent a realistic daytime link budget.

\begin{table}[htbp]
\centering
\caption{Numerical parameters used in the \gls{qkd} performance model.}
\label{tab:qkd-model-parameters}
\begin{tabular}{lll}
\hline
Symbol & Value & Description \\
\hline
$\mu$                    & 0.02                 & Average number of photon pairs per pump cycle \\
$D$                      & $100~\mathrm{s^{-1}}$  & Detector dark-count rate\\
$B$                      & $100~\mathrm{s^{-1}}$  & Background-count rate per \gls{ogs} \\
$N_{\mathrm{det}}$       & 2                    & Number of detectors per \gls{ogs} \\
$\tau_c$                 & 200~ps      & Coincidence time window \\
$e_0$                    & 0.5                  & Random-background error probability \\
$f_{\mathrm{EC}}$        & 1.22                 & Error-correction efficiency \\
$f_{\mathrm{rep}}$       & 1~GHz       & Pump repetition rate \\
$\varepsilon_{\mathrm{sec}}$       & $10^{-9}$       &  Secrecy criterion\\
$\varepsilon_{\mathrm{cor}}$       & $10^{-12}$       &  Correctness criterion
\\
$\beta$ & $\rightarrow \varepsilon_{\mathrm{sec}}/4$ & Auxiliary finite-key parameter (maximizes $\ell$) \\
\hline
\end{tabular}
\end{table}

\begin{table}[htbp]
\centering
\caption{\label{tab:qkd-performance-summary}
Pass-averaged $X$-basis \gls{qber} and asymptotic secret-key yield of orbit no.~9 for the four treatments. The pass-averaged \gls{qber} is weighted by the number of accepted coincidences, as in Eqs.~\eqref{equation:finite-key accepted events} and~\eqref{equation:finite-key erroneous events}. The $Z$-basis \gls{qber} is independent of the phase treatment, with $\bar{E}^Z \simeq 0.024\%$ and a maximum of 0.086\%.}
\begin{tabular}{lccc}
\hline
Treatment
& Phase information
& $\bar{E}^{X}$ (\%)
& Asymptotic key yield (bits) \\
\hline
1. Fixed parity
& none
& 50.1
& 72 \\
2. Relabeled parity
& $\Theta$ from ephemeris
& 18.1
& 144 \\
3. 1~Hz optical correction
& local phase (witness)
& 0.89
& 302 \\
4. Ideal optical correction
& local phase (exact)
& 0.024
& 325 \\
\hline
\end{tabular}
\end{table}

With ideal correction, the secret-key rate follows the combined dual-link transmittance (Fig.~\ref{fig:Valid Elevation And Slant Range}) and defines the envelope of Fig.~\ref{fig:secret-key-rate}. The 1~Hz sample-and-hold correction limits the maximum $X$-basis \gls{qber} to 2\% and maintains a positive rate throughout the common-access window. The difference from the ideal result arises from the residual phase accumulated during each 1~s hold interval, whereas the four isolated branch-selection errors shown in Fig.~\ref{fig:witness-phase-reconstruction} have a negligible impact on the pass-integrated yield. In the simulation, the phase-basis error is evaluated once at the end of each 1~s hold interval, when the residual phase is largest. As a result, the reported 1~Hz performance relies on an overestimate of the average error over each interval. 

Overall, the sample-and-hold correction preserves 93\% of the ideal pass-integrated asymptotic yield. Without optical correction, the secret-key rate periodically reaches zero for both fixed-parity and relabeled-parity treatments. With fixed parity, a key is generated only when $\mathcal{C}^X \simeq +1$, reducing the pass-integrated yield to 22\% of the ideal. Relabeling the parity recovers the extrema where $\mathcal{C}^X \simeq -1$ and doubles the yield to 44\%. Intervals near quadrature remain inaccessible because bit inversion compensates only a phase of $\pi$. Therefore, knowing the combined phase does not replace the need for optical correction.

To assess whether the positive asymptotic performance is sufficient for a finite data block, we evaluate the finite-key bound for orbit no.~9. We consider the ideal correction case, as it minimizes the phase-basis error and provides the most favorable bound. Equation~\eqref{equation:secret key length} gives an unclamped (i.e., not limited to the positive part) finite-key bound of $\ell \approx -156$ bits. Thus, although orbit no.~9 achieves a positive asymptotic secret-key rate, the finite number of accepted detections is still insufficient to generate a finite-length secret key, so $\max(0,\ell) = 0$. Even with ideal correction, the finite number of coincidence events collected in a single pass is insufficient to overcome finite-size statistical penalties~\cite{Staffieri_2026, Sidhu_2026}. We next investigate whether accumulating detection statistics over multiple common-access windows can overcome this finite-block limitation.

\subsection{Finite-key accumulation over successive passes}

Sidhu et al. showed that, in prepare-and-measure satellite \gls{qkd}, accumulating data from multiple satellite overpasses into a larger finite-key processing block can enable positive finite-length secret-key extraction, even when an individual overpass is insufficient~\cite{Sidhu_2022}. This approach, however, increases latency between data acquisition and secret-key generation~\cite{Sidhu_2022}. More recently, researchers have investigated finite-key performance for dual-downlink entanglement-based \gls{qkd}, including the effects of overpass geometry and finite coincidence statistics~\cite{Sidhu_2026}. These results also show that increasing the processing block size can reduce the finite secret-key length when the added data have substantially higher error rates~\cite{Sidhu_2026}.

For the considered ephemeris, the pass-averaged $X$-basis \gls{qber} ranges from 17.5\% to 18.9\% with relabeled parity, is about 50\% with fixed parity, and remains below 0.9\% with 1~Hz optical correction. The pass-averaged $Z$-basis and ideally corrected $X$-basis \glspl{qber} remain below $0.14\%$. We therefore accumulate the accepted detection and error counts over all successive common-access windows, taken in chronological order, and recalculate $\ell$ from Eq.~\eqref{equation:secret key length} using the resulting block-averaged \glspl{qber}. Within this ordering, orbit no.~9 is the second accumulated pass.

\begin{figure}[htbp]
\centering
\includegraphics[width=.7\columnwidth]{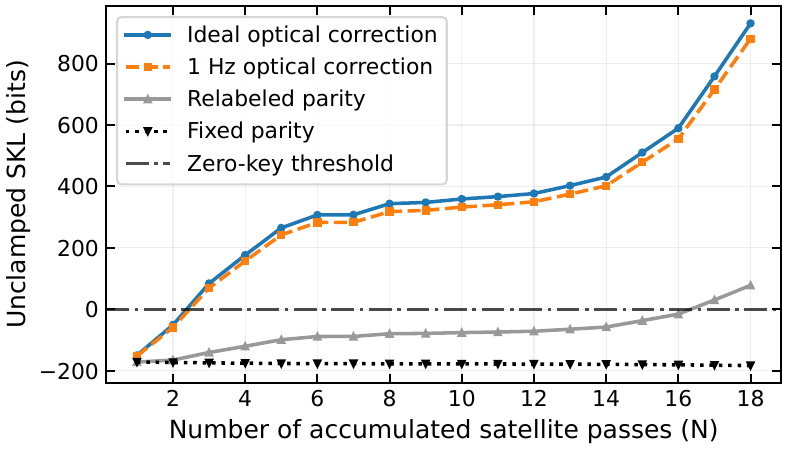}
\caption{Evolution of the unclamped finite secret-key length as successive common-access windows are accumulated in chronological order. The horizontal dash-dotted line indicates the zero-key threshold below which secret-key extraction is not yet possible. The finite secret-key length becomes positive after 3 accumulated passes for both ideal and 1~Hz correction, and after 17 passes with relabeled parity, whereas no positive finite key is obtained within the 18 passes with fixed parity.}
\label{fig:secret-key-length-evolution}
\end{figure}

Figure~\ref{fig:secret-key-length-evolution} shows the evolution of the finite secret-key length as detection statistics accumulate over successive passes. Because each additional pass contributes a different number of accepted detections due to its distinct link geometry, the finite secret-key length does not increase linearly with the number of accumulated passes. The ideal and 1~Hz sample-and-hold correction cases remain closely aligned and become positive after 3 accumulated passes. After 18 passes, the latter retains approximately 95\% of the ideal finite secret-key length, indicating that the residual phase error introduced by the sample-and-hold correction scheme has only a limited impact on finite-key performance. With relabeled parity, the finite-key bound remains below the zero-key threshold until 17 accumulated passes and reaches only about $8\%$ of the ideal finite secret-key length after 18 passes. With fixed parity, no positive finite key is obtained within the 18 passes, and since privacy amplification then consumes all the bits, each additional pass only increases the error-correction leakage and $\ell$ slowly decreases. The larger phase-basis error in the cases without optical correction increases the privacy-amplification cost and substantially reduces the secret key obtained from the accumulated detections.

\section{Discussion and limitations}
\label{section:Discussion and limitations}

The analysis identifies two practical implications for dual-downlink time-bin \gls{qkd}. First, after zero-Doppler calibration, the Doppler-induced carrier phase separates into independent signal- and idler-link contributions. As a result, each \gls{ogs} can track and correct its local contribution independently. Second, the proposed ephemeris-assisted witness-field scheme corrects the Doppler-induced carrier phase, but does not compensate for the detuning-dependent dephasing introduced by analyzer-delay mismatch. Both \glspl{ogs} must therefore set their analyzer delays to the nominal value $\tau_p$ before the pass and stabilize them around this value during operation. The zero-Doppler calibration of Eqs.~\eqref{equation:Local Calibration Condition} and~\eqref{equation:Pump Local Calibration Condition} is assumed to remain valid during each pass. This requires the pump frequency, time-bin separation, and analyzer phases to remain stable between calibration and operation.

The update rates in Table~\ref{tab:visibility-correction-rate} are evaluated for orbit no.~9, which exhibits the largest combined phase-evolution rate among the 18 common-access windows considered. As shown in Eq.~\eqref{equation:Doppler phase rate}, the required rate depends on the time-bin separation, carrier frequency, and line-of-sight acceleration. It increases with longer time-bin separations, shorter wavelengths, or lower orbits. The 1~Hz sample-and-hold correction rate adopted here therefore represents the considered orbital configuration rather than a general threshold for satellite time-bin \gls{qkd}.

Optical correction is also what makes the key extractable. Phase knowledge alone, even for both links combined, yields less than half the ideal asymptotic rate (Table~\ref{tab:qkd-performance-summary}) because only optical correction recovers intervals near quadrature. In the finite-key regime, reducing the $X$-basis error lowers the privacy-amplification cost and makes finite-size statistical fluctuations the main limitation as detections accumulate. Without optical correction, higher phase-basis error increases the required amount of data, and consequently the acquisition time for secret-key generation.

The finite-rate calculation of the witness-based correction preserves two deterministic effects. First, the phase accumulated during each 1~s hold interval creates a smooth difference between the corrected and ideal secret-key rates. Second, because the branch-selection metric uses finite differences between successive witness samples, a branch change near a witness contrast extremum may be delayed by one update. This produces isolated one-sample transients during which the residual phase can exceed the sample-and-hold bound of Eq.~\eqref{equation:combined-residual-phase-bound}. Beyond these effects, the witness phase reconstruction assumes noise-free witness measurements, accurate ephemeris data, correct initial fringe-order identification, continuous tracking, negligible actuation latency, and stable interferometers. The correction also requires a frequency-stable witness laser, since witness-frequency drift would be indistinguishable from the Doppler-induced phase evolution. In addition, the witness power must be sufficient to measure the contrast $\mathcal{C}_j$ with negligible shot-noise-limited phase error at the 1~Hz update rate. At the same time, its leakage into the quantum channel, set by the isolation of the wavelength demultiplexer, must remain small. These witness power and isolation requirements are left for future work. Moreover, like the ephemeris prediction, the proposed witness field architecture tracks only the link-dependent Doppler phase and does not account for pump-frequency drift or thermal and mechanical drift in the source and analyzer interferometers. Deriving the witness from the pump laser, or analyzing it in the same interferometer as the quantum signal, could extend its coverage to these drifts, but evaluating the witness under such conditions is beyond the scope of this work. Finally, the hardware implementation of the correction is not modeled: the insertion loss of the phase modulator and witness demultiplexer, the clock synchronization and time-slot tracking required to gate the modulator on the late time bin under Doppler-induced slot drift, and the polarization sensitivity of the modulator and receiver should be addressed before an experimental realization.

The \gls{qkd} model retains only vacuum and single-pair terms in the low-brightness regime and describes each channel by a total detection efficiency and a background-detection probability. The link calculation includes elevation-dependent atmospheric attenuation, diffraction-limited beam divergence, and fixed pointing loss, but neglects turbulence-induced beam broadening, scintillation, and time-dependent pointing jitter, which can be modeled following~\cite[Ch.~12]{Andrews_2005}. The security analysis assumes trusted, well-characterized measurement devices, so implementation attacks such as bright-light detector control in Franson-type systems~\cite{Jogenfors_2015} are not considered. The passive three-slot receiver relies on temporal post-selection, which opens the post-selection loophole in Franson-type interferometers. Bernardi et al. addressed this issue by implementing an integrated receiver that combines a fast electro-optic switch with an unbalanced interferometer. The switch directs the early and late time bins into the long and short arms, respectively, so that they overlap at the output, eliminating the need for post-selection~\cite{Bernardi_2026}. Extending the analysis to multipair emission~\cite{Ma_2007} and accounting for detector timing jitter, dead time, and afterpulsing would enable more realistic predictions of both single-pass and multi-pass \gls{qkd} performance.

Beyond the time-bin implementation considered here, the joint spectral-temporal description of the photon pairs, with the Doppler effect modeled as a transformation of the \gls{jsa} during propagation, applies directly to frequency-bin encoding and can be adapted to polarization encoding. The same description could be used to assess Doppler effects on Bell-inequality violation, measurement-device-independent \gls{qkd}, and entanglement swapping. However, the carrier-phase expressions and the correction scheme presented here assume \gls{umzi} receivers and would need to be re-derived for other measurement schemes, such as the switched receiver of Ref.~\cite{Bernardi_2026}.

\section{Conclusions}
\label{section:Conclusions}

We developed a spectral-temporal framework for time-bin-entangled photon pairs propagating through two satellite downlinks with distinct Doppler evolutions. For the considered low-Earth-orbit satellite pass and matched 400~ps analyzer delays, the received time-bin separation changes by at most 8.8~fs for each link. Yet, the combined carrier phase varies by approximately 38~rad, driving the interference through about six fringes and constituting the dominant Doppler contribution in the matched-delay configuration. 

After zero-Doppler calibration, the carrier phase separates into independent signal- and idler-link contributions. Each receiver can therefore correct its local phase without real-time phase exchange between the two \glspl{ogs},
whereas relabeling the parity requires the combined phase of both links. To compensate for this phase evolution, we proposed an ephemeris-assisted witness-field scheme in which the link-dependent Doppler-induced phase is reconstructed from the measured witness contrast, while ephemeris information resolves the phase-branch ambiguity. 

For orbit no.~9, the witness-derived correction limits the maximum simulated $X$-basis \gls{qber} to 2\% and retains 93\% of the ideal asymptotic pass-integrated key yield. However, the number of accepted detections collected during a single common-access window is insufficient for positive finite-key extraction, even under ideal carrier-phase correction. To overcome this finite-size limitation, we accumulated detections over successive passes. The finite secret-key length becomes positive after 3 accumulated passes for both ideal and 1~Hz sample-and-hold correction. After 18 passes, the 1~Hz correction retains approximately 95\% of the ideal finite secret-key length. Without optical correction, relabeling the parity from the ephemeris requires 17 passes to obtain a positive finite key and reaches only about 6\% of the ideal length after 18 passes, while a fixed parity yields no positive finite key within 18 passes.

Overall, this work establishes an end-to-end framework for satellite-based dual-downlink time-bin \gls{bbm92}, linking the Doppler-transformed biphoton state to receiver measurements and asymptotic and finite-key performance. It further shows that, with ideal phase estimation and precisely matched analyzer delays, the 1~Hz optical correction preserves near-ideal finite-key performance, whereas relabeling the parity from the ephemeris without optical correction does not. The witness-field measurement provides one way to obtain the required phase estimate locally at each \gls{ogs}.

\appendix

\section{Doppler-induced transformation of the time-bin biphoton state}
\label{appendix:Doppler-induced transformation of the time-bin biphoton state}

This appendix details the variable transformation from  the emitted state in Eq.~\eqref{equation:Emitted Time-Bin State} to the received state in Eq.~\eqref{equation:Received Time-Bin State}. Following Ref.~\cite{Schlake_2026, Schlake_2025_EPJQT}, the Doppler transformation for downlink $j \in \{s,i\}$ is represented by the frequency rescaling
\begin{equation}
\omega_j^r = \kappa_j\omega_j^e,
\qquad
\omega_j^e = \frac{\omega_j^r}{\kappa_j},
\qquad
d\omega_j^e = \frac{d\omega_j^r}{\kappa_j}.
\end{equation}
We assume the Doppler factor $\kappa_j$ remains constant during the biphoton coherence time and the early-late separation $\tau_p$. To preserve the continuous-frequency commutation relation, the receiver-frame creation operator must satisfy
\begin{equation}
\label{equation:appendix-operator-transformation}
a_j^{r\dagger}(\omega_j^r)
=
\frac{1}{\sqrt{\kappa_j}}
a_j^{e\dagger}
\left(
\frac{\omega_j^r}{\kappa_j}
\right),
\end{equation}
or, equivalently,
$a_j^{e\dagger}(\omega_j^e) = \sqrt{\kappa_j}a_j^{r\dagger}(\omega_j^r)$. Therefore, substituting Eq.~\eqref{equation:appendix-operator-transformation} into Eq.~\eqref{equation:Emitted Time-Bin State} yields
\begin{equation}
\label{equation:app-received-state-direct}
\begin{aligned}
\ket{\Psi^r}
&=
\frac{1}{\sqrt{2}}
\int d\omega_s^rd\omega_i^r
\frac{1}{\sqrt{\kappa_s\kappa_i}}
F^e
\left(
\frac{\omega_s^r}{\kappa_s},
\frac{\omega_i^r}{\kappa_i}
\right)
\left[
1 +
\exp\left(
i\left(
\frac{\omega_s^r}{\kappa_s}
+
\frac{\omega_i^r}{\kappa_i}
\right)\tau_p
+
i\phi_p
\right)
\right]\\
&\quad\times
a_s^{r\dagger}(\omega_s^r)
a_i^{r\dagger}(\omega_i^r)
\ket{0}.
\end{aligned}
\end{equation}
The factor $1/\sqrt{\kappa_s\kappa_i}$ ensures that the single-pulse joint spectral amplitude remains normalized
\begin{equation}
\label{equation:appendix-normalization}
\begin{aligned}
\int d\omega_s^r\,d\omega_i^r
\frac{1}{\kappa_s\kappa_i}
\left|
F^e
\left(
\frac{\omega_s^r}{\kappa_s},
\frac{\omega_i^r}{\kappa_i}
\right)
\right|^2
=
\int d\omega_s^e\,d\omega_i^e
\left|
F^e(\omega_s^e,\omega_i^e)
\right|^2
= 1.
\end{aligned}
\end{equation}
Using the Doppler phase defined in
Eq.~\eqref{equation:Doppler-induced phase mismatch},
\begin{equation}
\label{equation:app-phase-decomposition}
\left(
\frac{\omega_s^r}{\kappa_s}
+
\frac{\omega_i^r}{\kappa_i}
\right)\tau_p
=
(\omega_s^r + \omega_i^r)\tau_p
-
\sum_{j = s,i}
\phi_j^D(\omega_j^r).
\end{equation}
Substituting Eq.~\eqref{equation:app-phase-decomposition} into Eq.~\eqref{equation:app-received-state-direct} gives Eq.~\eqref{equation:Received Time-Bin State}.

\section{Analyzer path expansion and central-slot projection}
\label{appendix:Analyzer path expansion and central-slot projection}

This appendix presents the full post-analyzer pathway expansion used in Sec.~\ref{subsection:Interferometric analysis and BBM92 basis selection}, identifies the two overlapping amplitudes in the central coincidence slot, and derives the phase-basis coincidence probability. For a \gls{umzi} with beam-splitter convention $t = \sqrt{T}$, $r = i\sqrt{R}$, and $R + T = 1$, the field creation operator for a photon at frequency $\omega$ entering input port $A$ transforms as
\begin{equation}
\begin{aligned}
\label{appendix_equation:UMZI Mode Transformation}
a^\dagger(\omega)
&\rightarrow
i\sqrt{RT}
\left[
1 + e^{i(\omega\tau + \phi)}
\right]
c^\dagger(\omega)
+
\left[
T - R e^{i(\omega\tau + \phi)}
\right]
d^\dagger(\omega),
\end{aligned}
\end{equation}
Assuming $R = T = 1/2$ and omitting the overall phase $i$ at port $C$, the component exiting a selected port $\alpha \in \{C,D\}$ can be expressed as
\begin{equation}
\begin{aligned}
\label{equation:Port Dependent UMZI Mode Transformation}
a_j^\dagger(\omega_j)
&\rightarrow
\frac{1}{2}
\left[
1 + s_\alpha e^{i(\omega_j\tau_j + \phi_j)}
\right]
\alpha_j^\dagger(\omega_j)
\end{aligned}
\end{equation}
where $s_C = +1$ and $s_D = -1$, and $\alpha_j^\dagger$ creates a photon in the selected output port of analyzer $j$. Applying Eq.~\eqref{equation:Port Dependent UMZI Mode Transformation} independently to the signal and idler photons in Eq.~\eqref{equation:Received Time-Bin State} yields the unnormalized post-analyzer component for the output-port pair $\alpha,\beta \in \{C,D\}$
\begin{equation}
\label{equation:appendix-full-output-pair-component}
\begin{aligned}
\ket{\Psi^r_{\alpha\beta}}
&=
\frac{1}{4\sqrt{2}}
\int d\omega_s^rd\omega_i^r
\frac{1}{\sqrt{\kappa_s\kappa_i}}
F^e
\left(
\frac{\omega_s^r}{\kappa_s},
\frac{\omega_i^r}{\kappa_i}
\right)
\left[
1 +
\exp\left(
i(\omega_s^r+\omega_i^r)\tau_p
-
i\sum_{j=s,i}\phi_j^D(\omega_j^r)
+
i\phi_p
\right)
\right]\\
&\quad\times
\left[
1 + s_\alpha 
e^{
i(\omega_s^r\tau_s + \phi_s)
}
\right]
\left[
1 + s_\beta 
e^
{i(\omega_i^r\tau_i + \phi_i)}
\right]
\alpha_s^\dagger(\omega_s^r)
\beta_i^\dagger(\omega_i^r)
\ket{0}.
\end{aligned}
\end{equation}
The $1/4$ factor results from the two \gls{umzi} transformations, while the $1/\sqrt{2}$ comes from the emitted time-bin source state. Expanding Eq.~\eqref{equation:appendix-full-output-pair-component} yields the eight source-analyzer pathways for the zero-Doppler, matched-delay configuration, where $\kappa_s = \kappa_i = 1$ and $\tau_s = \tau_i = \tau_p$. Their positions in the receiver-frame coordinates are listed in Table~\ref{table:Path Combinations}. 

\begin{table}[htbp]
\centering
\caption{\label{table:Path Combinations}
Two-photon path combinations for matched delays assuming no-Doppler, ($\kappa_j = 1)$.}
\begin{tabular}{ccccc}
\hline
Path &
$t_s^r$ &
$t_i^r$ &
$t_+^r$ &
$t_-^r$ \\
\hline
$ESS$ &
$0$ &
$0$ &
$0$ &
$0$ \\
$ESL$ &
$0$ &
$\tau_p$ &
$\tau_p/2$ &
$-\tau_p$ \\
$ELS$ &
$\tau_p$ &
$0$ &
$\tau_p/2$ &
$\tau_p$ \\
$ELL$ &
$\tau_p$ &
$\tau_p$ &
$\tau_p$ &
$0$ \\
$LSS$ &
$\tau_p$ &
$\tau_p$ &
$\tau_p$ &
$0$ \\
$LSL$ &
$\tau_p$ &
$2\tau_p$ &
$3\tau_p/2$ &
$-\tau_p$ \\
$LLS$ &
$2\tau_p$ &
$\tau_p$ &
$3\tau_p/2$ &
$\tau_p$ \\
$LLL$ &
$2\tau_p$ &
$2\tau_p$ &
$2\tau_p$ &
$0$ \\
\hline
\end{tabular}
\end{table}

The $\ket{ELL}$ and $\ket{LSS}$ contributions coincide at $(t_+^r,t_-^r) = (\tau_p, 0)$ and form the central $X$-basis slot. The $\ket{ESS}$ and $\ket{LLL}$ pathways occur at $(0,0)$ and $(2\tau_p,0)$, respectively, and correspond to the early and late $Z$-basis outcomes.

Under ideal central-slot selection, retaining only the $\ket{ELL}$ and $\ket{LSS}$ contributions in Eq.~\eqref{equation:appendix-full-output-pair-component} yields
\begin{equation}
\begin{aligned}
\label{equation:appendix-full-central-slot-component}
\ket{\Psi^{r, X}_{\alpha\beta}}
&=
\frac{1}{4\sqrt{2}}
\int d\omega_s^rd\omega_i^r
\frac{1}{\sqrt{\kappa_s\kappa_i}}
F^e
\left(
\frac{\omega_s^r}{\kappa_s},
\frac{\omega_i^r}{\kappa_i}
\right)
\left[
e^{i\Phi_{\rm{LSS}}(\omega_s^r,\omega_i^r)} 
+
s_\alpha s_\beta
e^{i\Phi_{\rm{ELL}}(\omega_s^r,\omega_i^r)}
\right]\\
&\quad\times
\alpha_s^\dagger(\omega_s^r)
\beta_i^\dagger(\omega_i^r)
\ket{0},
\end{aligned}
\end{equation}
where
\begin{equation}
\begin{aligned}
\Phi_{\rm{ELL}}(\omega_s^r,\omega_i^r)
&=
\omega_s^r\tau_s +
\omega_i^r\tau_i + \phi_s + \phi_i\\
\Phi_{\rm{LSS}}(\omega_s^r,\omega_i^r)
&=
(\omega_s^r+\omega_i^r)\tau_p
-
\sum_{j = s,i}\phi_j^D(\omega_j^r)
+
\phi_p.
\end{aligned}
\end{equation}
Factoring the $LSS$ contribution from Eq.~\eqref{equation:appendix-full-central-slot-component} yields
\begin{equation}
\begin{aligned}
\label{equation:appendix-factored-central-slot-component}
\ket{\Psi^{r, X}_{\alpha\beta}}
&=
\frac{1}{4\sqrt{2}}
\int d\omega_s^rd\omega_i^r
\frac{1}{\sqrt{\kappa_s\kappa_i}}
F^e
\left(
\frac{\omega_s^r}{\kappa_s},
\frac{\omega_i^r}{\kappa_i}
\right) 
e^{i\Phi_{\rm{LSS}}(\omega_s^r,\omega_i^r)}
\left[
1 +
s_\alpha s_\beta 
e^{i\Phi(\omega_s^r, \omega_i^r)}
\right]\\
&\quad\times
\alpha_s^\dagger(\omega_s^r)
\beta_i^\dagger(\omega_i^r)
\ket{0},
\end{aligned}
\end{equation}
where the central-slot relative phase is
\begin{equation}
\label{equation:appendix-central-slot-relative-phase}
\begin{aligned}
\Phi(\omega_s^r,\omega_i^r)
&=
\Phi_{ELL}(\omega_s^r,\omega_i^r)
-
\Phi_{LSS}(\omega_s^r,\omega_i^r)\\
&=
\phi_s + \phi_i - \phi_p
+
\sum_{j \in \{s,i\}}
\omega_j^r
\left(
\tau_j - \frac{\tau_p}{\kappa_j}
\right).
\end{aligned}
\end{equation}
The Doppler-phase definition in Eq.~\eqref{equation:Doppler-induced phase mismatch} is applied in the second line. Equation~\eqref{equation:appendix-central-slot-relative-phase} represents the central-slot phase, as given in Eq.~\eqref{equation:Central slot phase}. The common factor $e^{i\Phi_{\rm{LSS}}(\omega_s^r,\omega_i^r)}$ is retained in the full state for calculating joint temporal distributions. Since this factor has unit modulus, it cancels out in the joint spectral intensity and frequency-integrated coincidence probabilities. Excluding it from the subsequent interference calculation yields the phase-reduced spectral amplitude defined in Eq.~\eqref{equation:phase-reduced central-slot state component}.

The central-slot coincidence probability for output-port pair $\alpha\beta$ equals the squared norm of the unnormalized component in Eq.~\eqref{equation:appendix-factored-central-slot-component}. Therefore, 
\begin{equation}
\label{equation:appendix-central-slot-probability-first-form}
\begin{aligned}
\mathcal{P}_{\alpha\beta}^{r,X}
=
\frac{1}{32}
\int d\omega_s^r d\omega_i^r 
\frac{1}{\kappa_s\kappa_i}
\left|
F^e
\left(
\frac{\omega_s^r}{\kappa_s},
\frac{\omega_i^r}{\kappa_i}
\right)
\right|^2
\left|
1 + s_\alpha s_\beta
e^{i\Phi(\omega_s^r,\omega_i^r)}
\right|^2.
\end{aligned}
\end{equation}
Using
\begin{equation}
\left|
1 + 
s_\alpha s_\beta 
e^{i\Phi(\omega_s^r, \omega_i^r)}
\right|^2
=
2\left[
1 + 
s_\alpha s_\beta 
\cos{
\left(
\Phi(\omega_s^r, \omega_i^r)
\right)}
\right],
\end{equation}
Eq.~\eqref{equation:appendix-central-slot-probability-first-form} becomes
\begin{equation} 
\begin{aligned} 
\label{appendix_equation:X-basis coincidence probability}
\mathcal{P}_{\alpha\beta}^{r, X}
&=
\frac{1}{16}
\int d\omega_s^rd\omega_i^r\frac{1}{\kappa_s\kappa_i} \Big|F^e
\Big(
\frac{\omega_s^r}{\kappa_s}, \frac{\omega_i^r}{\kappa_i} 
\Big)
\Big|^2
\Big[
1 + 
s_\alpha s_\beta 
\cos{
\big(
\Phi(\omega_s^r, \omega_i^r)
\big)}
\Big].
\end{aligned}
\end{equation}
By summing over the four output-port pairs and applying the transformed-\gls{jsa} normalization from Eq.~\eqref{equation:appendix-normalization}, we obtain the total coincidence probability in the $X$ basis
\begin{equation}
\label{equation:X-basis total coincidence probability}
p_X
=
\sum_{\alpha, \beta \in \{C, D\}}
\mathcal{P}_{\alpha\beta}^{r, X}
=
\frac{1}{4}
\int d\omega_s^rd\omega_i^r\frac{1}{\kappa_s\kappa_i} \Big|F^e
\Big(
\frac{\omega_s^r}{\kappa_s}, \frac{\omega_i^r}{\kappa_i} 
\Big)
\Big|^2
=
\frac{1}{4}.
\end{equation}
For equal source-bin amplitudes, with the analyzer differential delays matched to the source-bin separation,
\begin{equation}
\label{equation:Z-basis and mix total coincidence probability}
p_Z 
=
p_{\mathrm{ESS}} + p_{\mathrm{LLL}} 
=
\frac{1}{4},
\qquad
p_{\mathrm{mix}} 
=
p_{\mathrm{ESL}} + p_{\mathrm{ELS}} + p_{\mathrm{LSL}} + p_{\mathrm{LLS}}
=\frac{1}{2}.
\end{equation}
These probabilities are consistent with Singh et al.~\cite{Singh_2026}.

\section{Gaussian phase-basis visibility}
\label{appendix:Gaussian phase-basis visibility}

This section derives the phase-basis visibility using the central-slot coincidence probabilities derived in Appendix~\ref{appendix:Analyzer path expansion and central-slot projection}. The sum of the output-port-pair central-slot probabilities in Eq.~\eqref{equation:X-basis coincidence probability} over correlated and anticorrelated port pairs yields
\begin{equation}
\begin{aligned}
\label{equation:Correlated probabilities}
\mathcal{P}_{CC}^{r, X} 
+
\mathcal{P}_{DD}^{r, X}
&=
\frac{1}{8}
\int d\omega_s^rd\omega_i^r\frac{1}{\kappa_s\kappa_i} 
\left|F^e
\left(
\frac{\omega_s^r}{\kappa_s}, \frac{\omega_i^r}{\kappa_i} 
\right)
\right|^2
\left[1 + 
\cos{
\big(
\Phi(\omega_s^r, \omega_i^r)
\big)}
\right],
\end{aligned}
\end{equation}
and
\begin{equation}
\begin{aligned}
\label{equation:anticorrelated probabilities}
\mathcal{P}_{CD}^{r, X} 
+
\mathcal{P}_{DC}^{r, X}
&=
\frac{1}{8}
\int d\omega_s^rd\omega_i^r\frac{1}{\kappa_s\kappa_i} 
\left|F^e
\left(
\frac{\omega_s^r}{\kappa_s}, \frac{\omega_i^r}{\kappa_i} 
\right)
\right|^2
\left[1 -
\cos{
\big(
\Phi(\omega_s^r, \omega_i^r)
\big)}
\right].
\end{aligned}
\end{equation}
The Doppler transformation preserves the single-pair normalization 
\begin{equation}
\label{appendix:Doppler transformed normalization}
\int d\omega_s^rd\omega_i^r\frac{1}{\kappa_s\kappa_i} 
\left|F^e
\left(
\frac{\omega_s^r}{\kappa_s}, \frac{\omega_i^r}{\kappa_i} 
\right)
\right|^2 = 1.
\end{equation}
Substituting Eq.~\eqref{equation:Correlated probabilities} and Eq.~\eqref{equation:anticorrelated probabilities} into Eq.~\eqref{equation:Signed correlated contrast} therefore yields
\begin{equation}
\label{equation:Central slot signed contrast}
\mathcal{C}^X
=
\int d\omega_s^rd\omega_i^r\frac{1}{\kappa_s\kappa_i} 
\left|F^e
\left(
\frac{\omega_s^r}{\kappa_s}, \frac{\omega_i^r}{\kappa_i} 
\right)
\right|^2
\cos{
\left(
\Phi(\omega_s^r, \omega_i^r)
\right)}.
\end{equation}
Applying the carrier-detuning decomposition from Eq.~\eqref{equation:Central slot phase decomposition} and compensating the carrier contribution to ensure $\Theta = 0\pmod{2\pi}$, $\mathcal{V}^X = |\mathcal{C}^X|$ simplifies to
\begin{equation}
\mathcal{V}^X
=
\left|
\int d\Omega_s^r d\Omega_i^r
\frac{1}{\kappa_s\kappa_i}
\left|
F^e
\left(
\frac{\omega_{s0}^r + \Omega_s^r}{\kappa_s},
\frac{\omega_{i0}^r + \Omega_i^r}{\kappa_i}
\right)
\right|^2
\cos{
\left(
\Omega_s^rT_s
+
\Omega_i^rT_i
\right)
}
\right|.
\end{equation}
For the centered Gaussian source model with compensated carrier phase, evaluating this integral gives
\begin{equation}
\mathcal{V}^X
=
\exp\Bigg\{
-\frac{1}{2}
\left(\sigma_{\omega,s,\mathrm{marg}}^{r,I}\right)^2
\left[
T_s - \frac{\kappa_i}{\kappa_s}
\frac{(\sigma_{\omega,i}^I)^2}
{(\sigma_{\omega,p}^I)^2 + (\sigma_{\omega,i}^I)^2}T_i
\right]^2-
\frac{\kappa_i^2T_i^2}{2}
\frac{(\sigma_{\omega,p}^I)^2(\sigma_{\omega,i}^I)^2}{(\sigma_{\omega,p}^I)^2 + (\sigma_{\omega,i}^I)^2}
\Bigg\},
\label{supplement_equation:Exact Gaussian visibility}
\end{equation}
with the received marginal variance
\begin{equation}
\left(
\sigma_{\omega,s,\mathrm{marg}}^{r,I}
\right)^2
=
\kappa_s^2
\frac{
(\sigma_{\omega,s}^I)^2
\left[
(\sigma_{\omega,p}^I)^2
+
(\sigma_{\omega,i}^I)^2
\right]
}{(\sigma_{\omega,p}^I)^2
+
(\sigma_{\omega,s}^I)^2
+
(\sigma_{\omega,i}^I)^2}.
\label{appendix_equation:Signal marginal variance}
\end{equation}

\section{Witness-field phase reconstruction}
\label{appendix:Witness-field phase reconstruction}

This appendix presents the derivation of the classical phase observable and outlines the numerical phase-reconstruction method for receiver-side Doppler-phase monitoring. For each downlink $j\in\{s, i\}$, a bright continuous-wave witness field is transmitted toward an \gls{ogs} and analyzed using an auxiliary \gls{umzi}. The measured port contrast yields a wrapped estimate of the local Doppler-induced carrier phase. At each measurement interval, the appropriate inverse-sine branch is chosen based on the contrast rate and the ephemeris-derived Doppler-factor rate. The wrapped sequence is then unwrapped before conversion to the corresponding quantum wavelength. We write its emitted spectral amplitude as
\begin{equation}
\label{equation:app-witness-emitted-spectrum}
E_j^e(\omega^e)
=
E_{0}
G^e(\omega^e),
\end{equation}
where $E_0$ sets the optical power and $G^{\rm e}$ is a normalized spectral envelope common for each downlink, $\int d\omega^e \left|G^e(\omega^e)\right|^2 = 1$. Under the Doppler transformation $\omega^r = \kappa_j \omega^e$, the received spectral amplitude is
\begin{equation}
\label{equation:app-witness-received-spectrum}
E_j^r(\omega^r)
=
\frac{E_{0}}{\sqrt{\kappa_j}}
G^e\left(\frac{\omega^r}{\kappa_j}\right).
\end{equation}
The factor $1/\sqrt{\kappa_j}$ preserves the spectral-mode normalization under the change of variables between emitted and received frequencies.

\subsection{Unbalanced-interferometer response}
\label{appsubsec:Unbalanced-interferometer response}

The auxiliary witness interferometer introduces a delay $\tau_w$ and a phase $\phi_w$, adjusted during calibration and subsequently held fixed. For output ports $\alpha \in \{C, D\}$, with $s_C = + 1$ and $s_D = - 1$, the output field is
\begin{equation}
\label{equation:app-witness-output-C-spectrum}
E_{j\alpha}^r(\omega^r)
=
\frac{E_{0}}{2\sqrt{\kappa_j}}
G^e
\left(
\frac{\omega^r}{\kappa_j}
\right)
\left[
1 + s_{\alpha}
e^{i(\omega^r\tau_w + \phi_w)}
\right].
\end{equation}
Using
\begin{equation}
\left|
1 + s_{\alpha}e^{i(\omega^r\tau_w + \phi_w)}
\right|^2
=
2\left[
1 + s_{\alpha}\cos{(\omega^r\tau_w + \phi_w)}
\right],
\end{equation}
the output power is
\begin{equation}
\mathcal{P}_{j\alpha}(t)
=
\frac{|E_0|^2}{2}
\int d\omega^r
\frac{1}{\kappa_j}
\left|
G^e\left(\frac{\omega^r}{\kappa_j}\right)
\right|^2
\left[
1 + s_{\alpha}\cos{(\omega^r\tau_w + \phi_w)}
\right].
\end{equation}
We define the normalized output contrast as
\begin{equation}
\label{equation:app-witness-contrast}
\mathcal{C}_j
=
\frac{
\mathcal{P}_{jC} - \mathcal{P}_{jD}
}{
\mathcal{P}_{jC} + \mathcal{P}_{jD}
}.
\end{equation}
Given that 
\begin{equation}
\mathcal{P}_{jC} + \mathcal{P}_{jD}
=
\int d\omega^r
\frac{|E_0|^2}{\kappa_j}
\left|
G^e\left(\frac{\omega^r}{\kappa_j}\right)
\right|^2
= |E_0|^2,
\end{equation}
and
\begin{equation}
\mathcal{P}_{jC} - \mathcal{P}_{jD}
=
\int d\omega^r
\frac{|E_0|^2}{\kappa_j}
\left|
G^e\left(\frac{\omega^r}{\kappa_j}\right)
\right|^2
\cos{(\omega^r\tau_w + \phi_w)},
\end{equation}
the normalized contrast becomes
\begin{equation}
\label{equation:app-witness-contrast-2}
\mathcal{C}_j
=
\int d\omega^r
\frac{1}{\kappa_j}
\left|
G^e\left(\frac{\omega^r}{\kappa_j}\right)
\right|^2
\cos{(\omega^r\tau_w + \phi_w)}
\end{equation}
Equivalently, changing variables from the received frequency $\omega^r$ to the emitted frequency $\omega^e$, with $\omega^ r =\kappa_j\omega^e$ and
$d\omega^r = \kappa_j\,d\omega^e$, gives
\begin{equation}
\label{equation:app-witness-contrast-3}
\mathcal{C}_j
=
\int d\omega^e
\left|
G^e\left(\omega^e\right)
\right|^2
\cos{(\kappa_j \omega^e\tau_w + \phi_w)}.
\end{equation}

For a narrow-linewidth witness centered at the emitted carrier frequency $\omega_{w0}^e$, the phase variation across its spectral bandwidth is negligible. Using the normalization condition $\int d\omega^e |G^e(\omega^e)|^2 = 1$, the witness contrast reduces to $\mathcal{C}_j\simeq \cos\left(\kappa_j\omega_{w0}^e \tau_w + \phi_w\right)$. The witness \gls{umzi} is biased at quadrature in the zero-Doppler configuration, $\kappa_j = 1$, such that $\omega_{w0}^e\tau_w + \phi_w = \pi/2 \pmod{2\pi}$. The phase-to-contrast relation is therefore
\begin{equation}
\label{equation:witness-phase-contrast}
\mathcal{C}_j \simeq -\sin\left(\phi_{w,j}^D\right),\quad
\phi_{w,j}^D = \omega_{w0}^e \left(\kappa_j - 1\right)\tau_w.
\end{equation}
Thus, the auxiliary witness interferometer converts the local Doppler-induced carrier phase into a normalized output-power contrast.

\subsection{Ephemeris-assisted phase reconstruction} 
\label{appsubsec:Ephemeris-assisted phase reconstruction}

The contrast relation in Eq.~\eqref{equation:witness-phase-contrast} does not uniquely determine the Doppler-induced witness phase because two different phase values yield the same sine value ($\sin\theta = \sin(\pm \pi - \theta)$) within each $2\pi$ interval. For a witness contrast measurement at time $t_k$, the two possible inverse-sine solutions are
\begin{equation}
\begin{aligned}
\phi_{w,j}^{(1)}(t_k)
&=
- \arcsin\left(\mathcal{C}_j(t_k)\right),\\
\phi_{w,j}^{(2)}(t_k)
&=
\pi + \arcsin\left(\mathcal{C}_j(t_k)\right).
\end{aligned}
\label{equation:witness-phase-candidates}
\end{equation}
Here $\phi_{w,j}^{(1)} \in [-\pi/2,\pi/2]$ and $\phi_{w,j}^{(2)} \in [\pi/2,3\pi/2]$. The first candidate satisfies $\cos\phi_{w,j}^{(1)} \geq 0$, while the second satisfies $\cos\phi_{w,j}^{(2)} \leq 0$. Both candidates are consistent with the same contrast measurement, but only one represents the physical Doppler branch. We resolve this local branch ambiguity using the measured contrast evolution together with the Doppler factor derived from the orbital ephemeris. For two successive witness measurements, we evaluate
\begin{equation}
\dot{C}_j(t_k)
\simeq
\frac{\mathcal{C}_j(t_k) - \mathcal{C}_j(t_{k - 1})}{t_k - t_{k - 1}},
\label{equation:witness-contrast-rate}
\end{equation}
and
\begin{equation}
\dot{\kappa}_j(t_k)
\simeq
\frac{\kappa_j(t_k)-\kappa_j(t_{k - 1})}{t_k - t_{k - 1}}.
\label{equation:witness-kappa-rate}
\end{equation}
We define the branch-selection metric
\begin{equation}
\mathcal{B}_j(t_k)
=
\dot{\mathcal{C}}_j(t_k)
\dot{\kappa}_j(t_k).
\label{equation:witness-branch-metric}
\end{equation}

Differentiating Eq.~\eqref{equation:witness-phase-contrast} gives
\begin{equation}
\dot{\mathcal{C}}_j
=
-
\cos\left(\phi_{w,j}^D\right)
\dot{\phi}_{w,j}^D,\quad
\dot\phi_{w,j}^D
=
\omega_{w0}^e \tau_w \dot{\kappa}_j.
\end{equation}
Consequently
\begin{equation}
\label{equation:witness-branch-relation}
\mathcal{B}_j
=
-
\omega_{w0}^e\tau_w
\cos\left(\phi_{w,j}^D\right)
\dot\kappa_j^2.
\end{equation}

For $\mathcal{B}_j \neq 0$, the factor $\omega_{w0}^e \tau_w \dot\kappa_j^2$ is strictly positive, so the sign of $\mathcal{B}_j$ is opposite to that of $\cos(\phi_{w,j}^D)$. The physical inverse-sine branch is therefore selected according to
\begin{equation}
\widetilde{\phi}_{w,j}^D(t_k)
=
\begin{cases}
\phi_{w,j}^{(1)}(t_k), & \mathcal{B}_j(t_k) < 0,
\\[1mm]
\phi_{w,j}^{(2)}(t_k), & \mathcal{B}_j(t_k) > 0,
\\[1mm]
\tilde{\phi}_{w,j}^D(t_{k - 1}), & \mathcal{B}_j(t_k) = 0,\quad k > 1.
\end{cases}
\label{equation:witness-branch-selection}
\end{equation}

For $\mathcal{B}_j = 0$, the branch-selection metric does not identify either branch. This occurs either when $\cos(\phi_{w,j}^D) = 0$, or when $\dot\kappa_j = 0$. In the first case, the witness contrast is at an extremum of the interferometric response, the two inverse-sine candidates coincide modulo $2\pi$, so no branch distinction is required. In the second case, the Doppler evolution changes direction, the sign of $\mathcal{B}_j$ then provides no information for selecting between the two inverse-sine candidates. The numerical reconstruction therefore retains previously selected wrapped phase $\widetilde{\phi}_{w,j}^D(t_{k - 1})$ at $t_k$. If $\mathcal{B}_j = 0$ at the first reconstruction step, the wrapped phase remains undefined until a nonzero metric is obtained.

Although denoted by $\dot{C}_j(t_k)$ and $\dot{\kappa}_j(t_k)$, the finite-difference estimates in Eqs.~\eqref{equation:witness-contrast-rate} and~\eqref{equation:witness-kappa-rate} are calculated from the samples at $t_{k - 1}$ and $t_k$ and therefore represent the average evolution over the interval, not the instantaneous value at $t_k$. If the witness contrast crosses an extremum within the interval, the sign of $B_j(t_k)$ may still correspond to the branch before the extremum, causing the other inverse-sine candidate to be selected at $t_k$. In the present noise-free simulation, the correct branch is recovered at the next update, so each branch-selection error affects only one sample. These isolated one-sample transients are retained without filtering in the \gls{qkd} performance calculation.

\begin{figure}[htbp]
    \centering
    \includegraphics[width=.7\columnwidth]{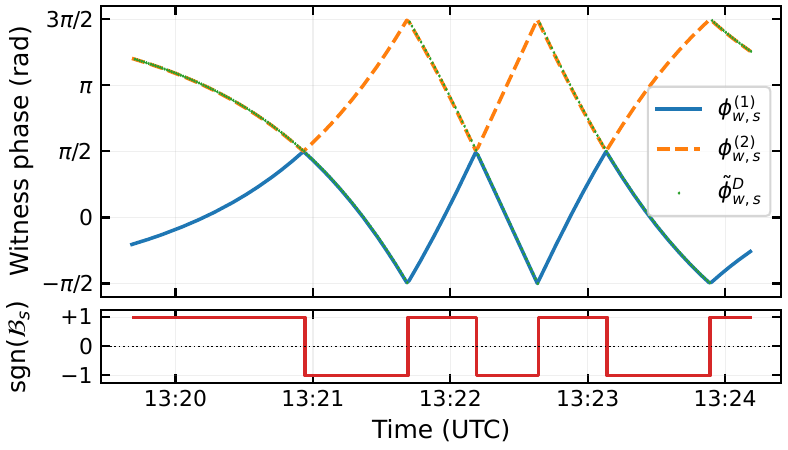}
    \caption{Ephemeris-assisted branch selection for the signal witness field. The upper panel displays the two inverse-sine candidates $\phi_{w,s}^{(1)}$ and $\phi_{w,s}^{(2)}$ obtained from the measured witness contrast, together with the selected wrapped phase $\tilde{\phi}_{w,s}^{D}$. The lower panel shows the branch-selection metric $\mathcal{B}_s = \dot{\mathcal{C}}_s \dot\kappa_s$.}
    \label{fig:witness-branch-selection}
\end{figure}

Figure~\ref{fig:witness-branch-selection}  illustrates the branch selection for the signal downlink. The reconstructed wrapped phase follows $\phi_{w,s}^{(1)}$ for $\mathcal{B}_s < 0$ and $\phi_{w,s}^{(2)}$ for $\mathcal{B}_s > 0$.

The branch-selected sequence $\tilde{\phi}_{w,j}^D(t_k)$ remains wrapped modulo $2\pi$. Temporal unwrapping removes successive $2\pi$ discontinuities and yields the continuous phase $\overline{\phi}_{w,j}^D(t_k)$. This determines the relative phase evolution but leaves the absolute fringe order undetermined. Using the ephemeris-predicted Doppler phase at the first reconstructed sample $t_1$, the integer fringe order is set once as
\begin{equation}
n_j
=
\operatorname{round}
\left(
\frac{
\omega_{w0}^e
\left(\kappa_j(t_1) - 1\right)\tau_w
-
\overline{\phi}_{w,j}^D(t_1)
}{2\pi}
\right),
\qquad
n_j \in \mathbb{Z}.
\label{equation:witness-fringe-order}
\end{equation}
The absolute reconstructed witness phase is then
\begin{equation}
\phi_{w,j,\mathrm{rec}}^D(t_k)
=
\overline{\phi}_{w,j}^D(t_k)
+
2\pi n_j.
\label{equation:witness-reconstructed-phase}
\end{equation}
The fringe order $n_j$ is kept fixed throughout the uninterrupted reconstruction interval.

\section*{Funding}
EIPHI Research Project  TACTICQ and QUANTICA (17-EURE-0002), ANR SPHIFA (ANR-20-CE47-0012), EQUIPEX+ Smartlight (21-ESRE-0040)

\section*{Acknowledgments}
The authors acknowledge the French National Research Agency (ANR) and the Bourgogne-Franche-Comté Region for their financial support.

\section*{Disclosures}
The authors declare no conflicts of interest.

\section*{Data availability}
The data supporting this paper’s results are not publicly available but can be provided by the authors upon reasonable request.

\FloatBarrier

\bibliography{main}

\end{document}